\documentclass[fleqn,usenatbib]{mnras}

\usepackage{newtxtext,newtxmath}

\usepackage[T1]{fontenc}
\usepackage[version=4]{mhchem}

\DeclareRobustCommand{\VAN}[3]{#2}
\let\VANthebibliography\thebibliography
\def\thebibliography{\DeclareRobustCommand{\VAN}[3]{##3}\VANthebibliography}

\usepackage{graphicx}	
\usepackage{amsmath}	
\usepackage{gensymb}
\usepackage{caption}
\usepackage{subcaption}
\usepackage{multirow}
\usepackage{tabularx}
\usepackage{titlesec}
\usepackage{csquotes}
\usepackage{xcolor}
\usepackage{mathtools}
\usepackage{multicol}
\usepackage{float}

\usepackage{footnote}
\makesavenoteenv{tabular}
\usepackage{tablefootnote}
\usepackage{threeparttable}
\usepackage{orcidlink}

\graphicspath{{plots/},{../plots/}}

\titlespacing{\section}{0pt}{0.75cm}{0.5cm}
\titlespacing{\subsection}{0pt}{0.5cm}{0.25cm}

\titleformat{\section}{\normalfont\fontsize{12}{15}\bfseries}{\thesection}{1em}{}
\def\msun{ \mathrm{M}_{\odot}}
\def\kms{\mathrm{km}\,\mathrm{s}^{-1}}

\title[Mock NIRSpec Observational Galaxy Kinematics]{Lessons Learned from Studying H$\alpha$ Galaxy Kinematics with Mock JWST/NIRSpec IFU Observations at $z > 6$}

\author[S. G. Phillips et al.]{
Si{\^a}n Phillips$^{1}$ \thanks{E-mail: S.G.Phillips@2017.ljmu.ac.uk}\orcidlink{0000-0003-4652-1090},
Francesca Rizzo$^{2,3,4}$ \orcidlink{0000-0001-9705-2461},
Mahsa Kohandel$^{5}$ \orcidlink{0000-0003-1041-7865},
Renske Smit$^{1}$ \orcidlink{0000-0001-8034-7802},
Andrea Pallottini$^{5,6}$ \orcidlink{0000-0002-7129-5761}
\\
$^{1}$ Astrophysics Research Institute, Liverpool John Moores University, 146 Brownlow Hill, Liverpool L3 5RF, UK\\
$^{2}$ Kapteyn Astronomical Institute, University of Groningen, Landleven 12, 9747 AD Groningen, The Netherlands \\
$^{3}$ Cosmic Dawn Centre (DAWN)\\
$^{4}$ Niels Bohr Institute, University of Copenhagen, Jagtvej 128, 2200 Copenhagen N, Denmark\\
$^{5}$ Scuola Normale Superiore, Piazza dei Cavalieri 7, I-56126 Pisa, Italy\\
$^{6}$ Dipartimento di Fisica ``Enrico Fermi'', Universit\'{a} di Pisa, Largo Bruno Pontecorvo 3, Pisa I-56127, Italy
}

\date{Accepted XXX. Received YYY; in original form ZZZ}

\pubyear{2025}

\begin{document}
\label{firstpage}
\pagerange{\pageref{firstpage}--\pageref{lastpage}}
\maketitle

\begin{abstract}
Galaxies with a disk morphology have been established at $z > 9$ with the James Webb Space Telescope (JWST). However, confirming their disky nature requires studying their gas kinematics, which can be challenging when relying solely on the warm gas observed by JWST. Unlike the cold gas traced by the Atacama Large Millimetre/Submillimetre Array (ALMA), warm gas is sensitive to outflows, complicating the interpretation of the disk dynamics. This elicits the question of how to compare information obtained from varied tracers, as well as how to physically interpret the low angular and spectral resolution observations generally available at high redshift. We address these challenges through comparative kinematic analysis of idealised and realistic NIRSpec/IFU mock observations derived from two galaxies in the SERRA suite of cosmological zoom-in simulations. With these synthetic data, we determine the robustness of dynamical information recovered from typical IFU observations, and test widely-used criteria for identifying disks and gaseous outflows at high redshift. We find that at the typical NIRSpec/IFU spectral and angular resolution ($\sim$ 0.05"/pixel), non-circular motions due to inflows or outflows can mimic the smooth velocity gradient indicative of a disk, and bias measured velocity dispersion upwards by a factor of $2-3\times$. As a result, the level of rotational support may be underestimated in the NIRSpec/IFU observations. However, the recovered dynamical mass appears to be relatively robust despite biases in $v_\text{rot}$ and $\sigma$.  
\end{abstract}




\section{Introduction}
\noindent The evolution of galaxies is governed by a complex interplay of astrophysical phenomena including minor and major merger events, smooth gas accretion, and gaseous outflows driven by feedback from stars and active galactic nuclei \citep[e.g.][]{Dayal18, Crain23}. These processes leave distinct dynamical signatures in a galaxy, and therefore the study of galaxy kinematics represents a unique probe of the mechanisms governing the mass assembly and growth of galaxies. 

The James Webb Space Telescope (JWST) is opening up an unprecedented view into the first galaxy populations, allowing the study of both their stellar morphology through the analysis of rest-frame near-infrared or optical imaging, and their kinematics through stars \citep[e.g.][]{DEugenio24}, and emission lines (e.g., H$\alpha$, [O{\sc iii}]) tracing the warm ionized gas. Early morphological analysis of JWST imaging suggested that the disk population at $z >$ 1.5 may be up to $10 \times$ greater than was seen by the Hubble Space Telescope \citep{Ferreira22}. Disks have been discovered up to $z \sim 9$ in the Cosmic Evolution Early Release Science Survey \citep[CEERS,][]{Ceers23} using morphological criteria, whereby galaxies are classified as potential disks if they display flattened, extended light distributions \citep[e.g.][]{Robertson23, VegaFerrero24}. The disk fraction estimated from a large sample of CEERS galaxies is 60\% at $z = 3$ and $\sim$ 30\% at $z = 6-9$ \citep{Kartaltepe23}.
Kinematic information is crucial for positively confirming that visually identified disk candidates are indeed rotational systems \citep[e.g.][]{Wisnioski15, Simons19, Rizzo22, Wang24}. One widely employed metric used to quantify the degree of rotational support in a galaxy is the ratio of rotational velocity to velocity dispersion, v/$\sigma$, where the velocity dispersion measures the turbulence of the ISM through the broadening of the spectral line \citep{Li23, Nelson23, deGraaff24}. A value of v/$\sigma \geq 3$ is considered an indicator of a settled disk \citep{ForsterSchreiber20}, and v/$\sigma \geq 10$, typical of local galaxies, indicates a dynamically cold disk. 

Galaxy kinematic studies using warm ionized gas tracers, most notably H$\alpha$, at intermediate redshifts ($ 0.5 \lesssim z \lesssim 3.5 $) have found that the velocity dispersion within galaxies increases, and the v/$\sigma$ ratio decreases, as a function of redshift \citep{Wisnioski15, Turner17, Johnson18, Ubler19, Birkin24} such that galaxies at $z \geq 2.5$ have v/$\sigma < 2$ \citep[e.g.][]{Wisnioski15, Turner17}.
This is in agreement with simulations \citep[e.g.][]{Ceverino17, Hung19}. For instance, using TNG50 simulations, \citet{Pillepich19} find that at $z < 5$, disks traced by H$\alpha$ become significantly more turbulent with increasing redshift.

Cold gas observations of galaxies, especially [C {\sc ii}]-158$\mu$m and CO observations with the Atacama Large Millimetre/Submillimetre Array (ALMA), are characterised by lower velocity dispersion and a greater degree of rotational support (i.e., $v/\sigma \sim 10$)  at $z = 0.5-6$ \citep[][ but see \citealp{Spilker22}]{Jones17, Neeleman20, Rizzo20, Rizzo21, Fraternali21, Tsukui21, Lelli21, Lelli23, Rizzo23, R-O23}. Marginally resolved kinematic characterisations of sources in the Epoch of Reionization (EoR) indicate that disk structure may already be present at these early times \citep{Smit18, Posses23, Parlanti23}, while the high resolution observations presented in \cite{Rowland24} reveal the earliest dynamically cold disk yet discovered, with v/$\sigma \sim 10$ at $z = 7.31.$ 

\cite{Rizzo24} examine the discrepancy between warm and cold gas dynamics by comparing the velocity dispersion and v/$\sigma$ measurements using spatially resolved CO, [C{\sc i}] and [C{\sc ii}] observations from the ALPAKA sample at $z = 0.5-3.5$ \citep{Rizzo23} and from the literature \citep{Girard19, Girard21, Bacchini20, Fraternali21, Rizzo21, Lelli21, Lelli23, R-O23} with trends from observations and models of warm ionized gas. They find that H$\alpha$ yields measurements of $\sigma$ (v/$\sigma$) that are higher (lower) than those from cold gas by a factor of $\sim$ 3. 

This observational result is in accord with theoretical work by \citet{Kohandel24}, using the SERRA zoom-in cosmological simulations \citep{Pallottini22}. \citet{Kohandel24} compare simulated velocity dispersion and v/$\sigma$ values obtained from H$\alpha$ and [C{\sc ii}] at $z = 4 - 9$, finding that the velocity dispersion from [C{\sc ii}] is a factor of $2 - 3 \times$ smaller, on average, than the H$\alpha$ value.
They find that v/$\sigma$ does not strongly evolve with redshift, and show that the tracers probe different galactic regions. [C{\sc ii}] traces the disk, as [C{\sc ii}] originates from the cold neutral medium and is found around molecular clouds \citep{Vallini15, Pallottini17feedback, Olsen21}, while H$\alpha$ traces both the disk and surrounding ionized gas, such as outflowing or inflowing streams \citep[see also][]{Ejdetjarn22, Ejdetjarn24}.
This contamination of warm gas kinematics by the non-circular motion of gas surrounding the galaxy is primarily responsible for the difference between kinematics as measured by cold and warm gas tracers. 

Hence, cold gas observations provide a less biased estimate of the disk velocity dispersion, rotational velocity and circular velocity. To benefit from the synergy between JWST and ALMA, it is imperative to be able to interpret the difference in kinematic measurements from warm and cold gas. 
However, as discussed in \cite{Rizzo22}, it is challenging to obtain high resolution and SNR observations at z $>$ 4 with ALMA, the primary facility currently available for observing galaxies in cold gas. Existing ALMA high-resolution surveys of galaxies at $z > 4$ are therefore biased towards bright, massive galaxies with high star formation rates \citep[e.g.][]{LeFevre20, Bouwens22, Li24}. The JWST/IFU allows some redress of this bias towards extreme sources at $z > 4$, as it is capable of observing normal galaxies from this redshift range in warm gas, and thus greatly extends the accessible redshift range for warm gas kinematics. Recent studies analyze the kinematics of individual objects in the EoR at
high resolution with the G395H grating (R $\sim$ 2700). They identify a galaxy group with indications of merger activity in its constituent galaxies at $z = 6.34$ \citep{Jones24}, a candidate unsettled, turbulent disk at $z=6.9$ \citep{Arribas24}, a galaxy at $z=6.9$ interpreted as a merging system \citep{Scholtz25}, and a lensed galaxy with a velocity gradient that could be indicative of rotation or merger activity at $z = 9.11$ \citep{Marconcini24}. 
In the context of such progress, the aim of this paper is to understand whether we can robustly identify galaxies hosting disks at $z > 6$ and derive their intrinsic properties, including the level of turbulence and the presence of outflows, with the typical observations available from the JWST/NIRSpec IFU. 
To achieve this, we create idealised and realistic mock NIRSpec observations for two galaxies from the {\sc serra} simulations \citep{Pallottini22}, which form part of the sample studied in \cite{Kohandel24}. These are representative massive disk galaxies ($\sim 10^{10}$M$_{\odot}$) at redshift $z=6-7$, chosen based on the comparison between the kinematics as measured by H$\alpha$ and [C{\sc ii}] (see Section \ref{targets}). One appears to host strong outflows, while the other shows no sign of non-circular motion. They act as case studies for which we examine the dynamical properties that can be recovered from observations using state-of-the-art analysis techniques.
The structure of the paper is as follows: in Section \ref{serra} we introduce the suite of simulations our target galaxies are drawn from, and establish their properties. In Section \ref{Creating_Obs} we outline the process by which both idealised and realistic mock NIRSpec observations are created from the simulation outputs. In Section \ref{Techniques} we describe the methods we use to analyse the data. The results are discussed in Section \ref{results_section} and summarised in Section \ref{Conclusions}. 

\section{SERRA Simulations}\label{serra}

\noindent {\sc serra} is a suite of high resolution ($\simeq 10^4$M$_{\odot}$, $\simeq 20\,\mathrm{pc}$) zoom-in cosmological simulations which tracks the formation and evolution of galaxies at $z > 4$ \citep{Pallottini22}. These simulations model the interactions between radiation, gas, stars, and dark matter within a cosmological framework, using the adaptive mesh refinement code {\sc ramses} \citep{Teyssier02, Rosdahl13}. Combined with the zoom-in technique, this provides the spatial and temporal resolution necessary for studying small-scale processes, including star formation and feedback \citep[e.g. from SN and stellar winds:][]{Pallottini17feedback}, chemical evolution \citep[non-equilibrium chemistry up to molecular hydrogen formation:][]{Grassi14, Pallottini17chemistry}, and radiative effects \citep[e.g. photoionization and photoevaporation:][]{decataldo19, Pallottini19}.

In post-processing \citep[cfr.][]{lupi:2020, katz:2025}, the {\sc serra} simulations can be used to analyze line emission across multiple wavelengths using realistic, observational-like pipelines \citep{zanella:2021, Rizzo22}, to provide detailed insights into the physical conditions of high-redshift galaxies. 
[C{\sc ii}] and H$\alpha$ line emission in {\sc serra} is computed using the {\sc cloudy} spectral synthesis code \citep{Ferland17}, accounting for the turbulent structure of molecular clouds \citep{vallini18, Pallottini19, Kohandel20}. For each emission line, simulated integral field unit observations are created from the output of the zoom-in simulations in the form of hyperspectral data cubes \citep{Kohandel19, Kohandel20}. 
These are cubic regions with two spatial and one spectral dimension, centred on target galaxies, upon which the velocity-dependent line surface brightness is modeled using Gaussian kernels to account for ISM temperature and internal turbulence.

\begin{figure*}
	\includegraphics[width=\textwidth]{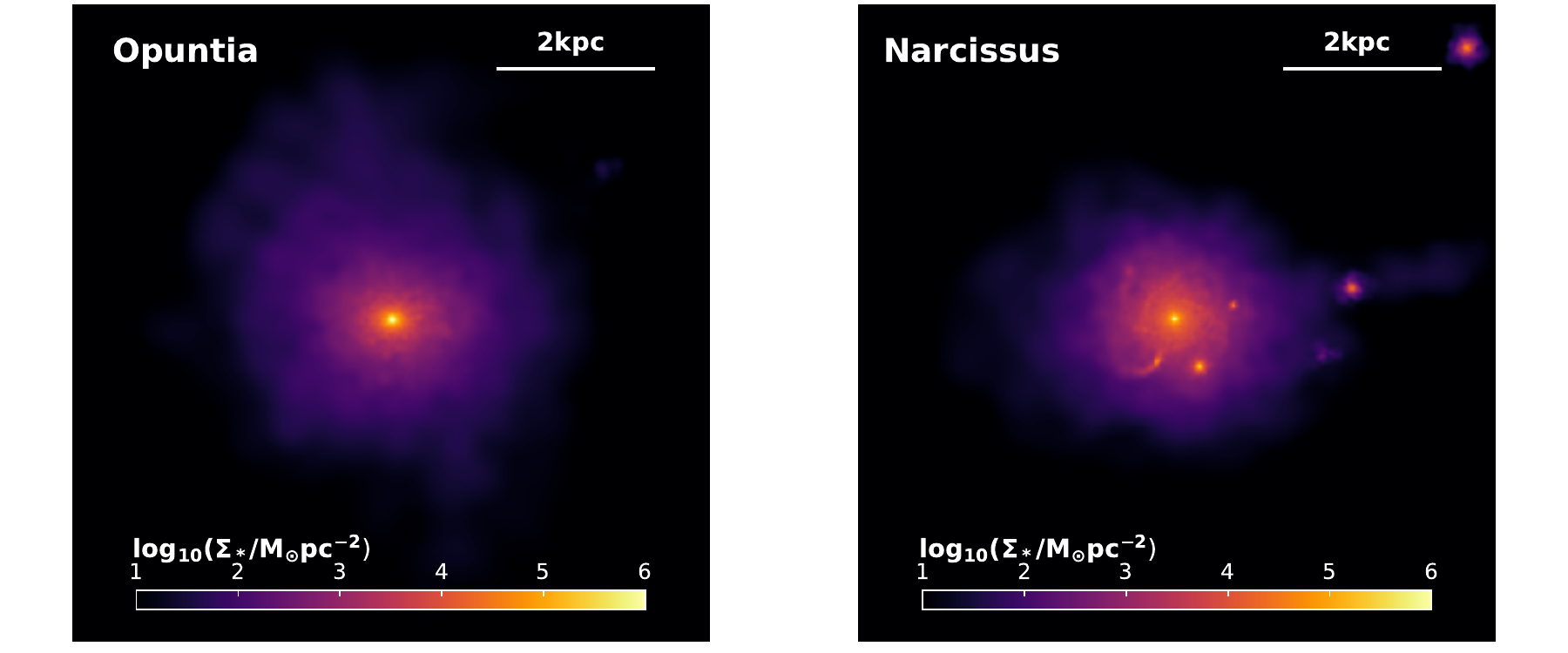}
    \caption{Maps of stellar distribution, with a colorbar showing the stellar mass density per unit area. The field of view is 8kpc $\times$ 8kpc, corresponding to $\sim$ 1.4" $\times$ 1.4" at $z = 6.07$ (Opuntia, $M_\star = 1.2 \times 10^{10} \msun$) and $\sim$ 1.5" $\times$ 1.5" at $z = 6.82$ (Narcissus, $M_\star = 1.0 \times 10^{10} \msun$). }
    \label{fig:stars}
\end{figure*}

\subsection{Target Sources}
\label{targets}

\begin{table*}
	\centering
	\caption{Properties of the {\sc serra} galaxies derived directly from the simulations. }
	\label{tab:target_properties}
	\begin{tabular}{|c|cccccccccccc|} 
		\hline
		Name & $z$ & M$_\star$  & M$_\text{gas}$  &  L$_{\mathrm{H}\alpha}$ &  L$_\mathrm{[CII]}$  & SFR   &  $r_{\mathrm{eff}, \mathrm{H}\alpha}$   &    $r_{\mathrm{eff}, \mathrm{[CII]}}$      & $\sigma_{[\text{C{\sc ii}}]}$ & $\sigma_{\text{H}\alpha}$ & (v/$\sigma)_{[\text{C{\sc ii}}]}$ & (v/$\sigma)_{\text{H}\alpha}$\\
  		    &      &  [$10^{10}\msun$] & [$10^{9}\msun$] & [$10^9$L$_{\odot}$] & [$10^9$L$_{\odot}$] & [M$_{\odot}$yr$^{-1}$] & [kpc] & [kpc] &  [$\kms$]                     & [$\kms$]                  &                                   &  \\
		\hline
		Opuntia & 6.07 & 1.2 & 4.5 &  1.8 &  0.3 & 19 &  1 &  0.7 & 26.8 & 58.6 & 8.8 & 4.1 \\
		Narcissus & 6.82 & 1.0 & 4.1 &  23 &  1.1 & 38.4 &  0.8 &  0.8 & 20.8 & 23.9 & 11.9 & 10.4 \\
		\hline
	\end{tabular}
\end{table*}

\noindent In this paper, we consider two simulated star-forming galaxies from the {\sc serra} suite as representative case studies: \enquote{Opuntia} at $z = 6.1$ (SFR $= 19$ M$_{\odot}$yr$^{-1}$, $M_{\star} = 1.2 \times 10^{10}\msun$) and \enquote{Narcissus} at $z = 6.8$ (SFR $= 38.4$ M$_{\odot}$yr$^{-1}$, $M_\star = 1.0 \times 10^{10}\msun$). 

These galaxies were selected from the \citealt{Kohandel24} sample ($\simeq 3000$ galaxies at $4\leq z \leq 9$) based on their kinematic properties, as as each hosts a dynamically cold disk. From this sample, the authors find that 60\% of sources are dynamically cold in [C{\sc ii}].  \\
We selected our two target galaxies based on their contrasting multi-wavelength kinematic properties. Narcissus, introduced in \cite{Kohandel24}, is an EoR galaxy that maintained a cold disk for over ten dynamical times, showing consistent $v/\sigma$ in both [C{\sc ii}], a tracer of cold gas, and H$\alpha$, which traces the warm ionized component. The second galaxy, Opuntia, was previously analysed in \cite{Rizzo22} and displays a significantly warmer ionized gas component (see intrinsic\footnote{In \citet{Kohandel24}, the intrinsic $\sigma_{\mathrm{em-line}}$ is defined as the luminosity-weighted average velocity dispersion, calculated using moment-2 and moment-0 maps of the respective emission line. The rotational velocity $V_{\mathrm{rot}}$ is estimated using the galaxy’s circular velocity, given by $V_{c} = (GM_{\mathrm{dyn}}/r_d)^{1/2}$ where $M_{\mathrm{dyn}} = M_{*} + M_{g}$ represents the dynamical mass within the desired field of view, and $r_d$ is the disk effective radius, defined as the radius containing 50$\%$ of the gas mass.} values in Table \ref{tab:target_properties}). This contrast in the cold and warm gas kinematics of high-redshift galaxies has been found to be a signature of outflows \citep[see][]{Kohandel24, Kohandel25}.

To give an overview of the galaxies, in Figure \ref{fig:stars} we show their stellar distributions. The stellar mass is strongly centrally concentrated in Opuntia, while Narcissus additionally hosts pockets of high stellar density. The most relevant global properties for the galaxies are recorded in Table \ref{tab:target_properties}. Narcissus has a high H$\alpha$ luminosity, leading to an offset relative to widely-used L$_{\text{H}\alpha}$-SFR calibrations such as the one presented in \cite{Kennicutt98}. This offset is not unexpected for galaxies in the EoR, as the \cite{Kennicutt98} relation assumes solar metallicity and continuous star formation over 100 Myr. In contrast, {\sc serra} galaxies form stars in relatively metal-poor bursty episodes \citep{Pallottini25}, which can enhance the ionizing photon production per unit SFR and thus increase the predicted H$\alpha$ luminosity (Kohandel et al., in prep). Similar excesses have been reported in other zoom-in simulations (e.g. SPHINX, \citealt{Katz19}), where low metallicity and bursty star formation systematically raise line luminosities. \\
As shown in \cite{Pallottini22}, the L$_\text{[CII]}$-SFR relation of {\sc serra} galaxies is consistent with $z\sim 4$ observations \citep[e.g.][]{Carniani18}, and similar to the relations at $z=0$ for starburst galaxies \citep{deLooze14, HC18} albeit with a larger scatter (0.48 dex), which can cause individual galaxies to appear above (Opuntia) or below (Narcissus) the average of the relation.

\section{Creating Mock Observations}\label{Creating_Obs}

\noindent Our ability to reliably derive disk kinematic properties in the presence of significant non-circular motions is limited by the spectral resolution, angular resolution, and signal to noise ratio.
In order to test the effect of varying SNR and angular resolution on the feasibility of recovering the main properties of disks at $z > 4$ and the presence of non-circular motions driven by outflows, we create both realistic mock NIRSpec and idealised observations from the two simulated galaxies. The simulated datacubes have a native spectral pixel size of 10.1 $\kms$ and spatial pixel size of $\sim 0.03$ kpc, corresponding to an angular resolution of $\sim 0.005$~arcsec at the redshifts of our targets. The field of view is 8 $\times$ 8 kpc ($\sim$ 1.4" $\times$ 1.4" for Opuntia, $\sim$ 1.5" $\times$ 1.5" for Narcissus).
We use the simulation output to create idealised and mock NIRSpec observations using the general formula:
\begin{equation}\label{eq:mock_observations}
   \textrm{ processed cube = (simulated cube} \otimes \textrm{PSF} \otimes \textrm{LSF}) + \textrm{noise}\,,
\end{equation}
where $\otimes$ represents a convolution operator, PSF is the Point Spread Function, and LSF is the Line Spread Function. Different PSFs, LSFs, and noise approximations are used for the idealised (Sec. \ref{creating_ideal}) and realistic (Sec. \ref{creating_realistic}) cases.

\subsection{Idealised Data}\label{creating_ideal}

\noindent We do not perform any spatial or spectral rebinning, in order to preserve the high intrinsic resolution in the creation of idealised observations, and so permit a comparison of the information recoverable from realistic mock NIRSpec observations and observations with ideal data quality. We convolve with a small PSF so as to maximise the number of resolution elements sampling the galaxy and therefore test the impact of resolution on our ability to recover kinematic properties from H$\alpha$. The process of creating the idealised data is outlined in the following.
\begin{enumerate}
    \item The pixel size ($\sim 0.005"$) and the spectral pixel size (10.1 $\kms$) are not changed from the simulation output.
    \item The data were convolved with a PSF and LSF (eq. \ref{eq:mock_observations}). The PSF is formed from a 2D Gaussian with Full Width at Half Maximum (FWHM) of 3 pixels $\approx 0.016"$. The LSF has a FWHM equivalent to twice the spectral pixel size.
    \item We add noise taking a realization from a normal distribution, setting a high S/N ratio of 50 in an aperture of equivalent size to the PSF at the outer extent of the galaxy, so as to maximise signal while still providing a noise background to facilitate kinematic fitting. 
\end{enumerate}

\subsection{Realistic mock NIRSpec Data}\label{creating_realistic}
\noindent The process of creating the mock observations was to interpolate the data to the spectral and angular pixel scale of NIRSpec, and then to add a realistic realisation of noise. 
\noindent The highest spectral resolution mode of JWST/NIRSpec, designed for kinematic characterisation of distant galaxies, has a resolution $R \simeq 2700$ \citep{Jakobsen22}, which may not be sufficient to accurately measure velocity dispersion values below $\sim 50$ $\kms$, as a single spectral resolution element has a width of $\approx 111/\sqrt{8ln(2)} = $ 47.1 $\kms$, and the instrumental LSF may here introduce strong uncertainties (e.g. \citealt{Lelli23}, and see \citealt{deGraaff24} in which the LSF of the NIRSpec dispersers is found to be a strong function of the target light profile, with the measured LSF being up to $2\times$ lower than reported by STSci - see their Appendix A.)  The pixel scale of JWST is $\sim$ 0.1"/pixel, equating to $\sim$ 0.6 kpc/pixel at $z = 6$, and in IFU mode the FoV is $\sim 3.0" \times 3.0"$ \citep{Nirspec22III}. 

To simulate detector noise we ran the JWST Exposure Time Calculator \citep[ETC:][]{Pontoppidan16} on a blank scene, using the grating G395H, which provides the best spectral resolution for H$\alpha$ at our target redshifts. The readout pattern employed is NRSIRS2, which offers an improvement in the handling of correlated noise compared to traditional readout methods through regular sampling of reference pixels, and is particularly effective for long exposures of faint sources \citep{Rauscher2012, Birkmann22}. The detector parameters are informed by the set-up utilised to perform integral field spectroscopy on galaxies at $z > 8$ in GTO Cycle 1 Programme 1262 (PI Luetzgendorf). The specific details of creating the idealised and mock observations are described in the following.

\begin{figure*}
     \centering
     \begin{subfigure}{\textwidth}
         \centering
         \includegraphics[width=\textwidth]{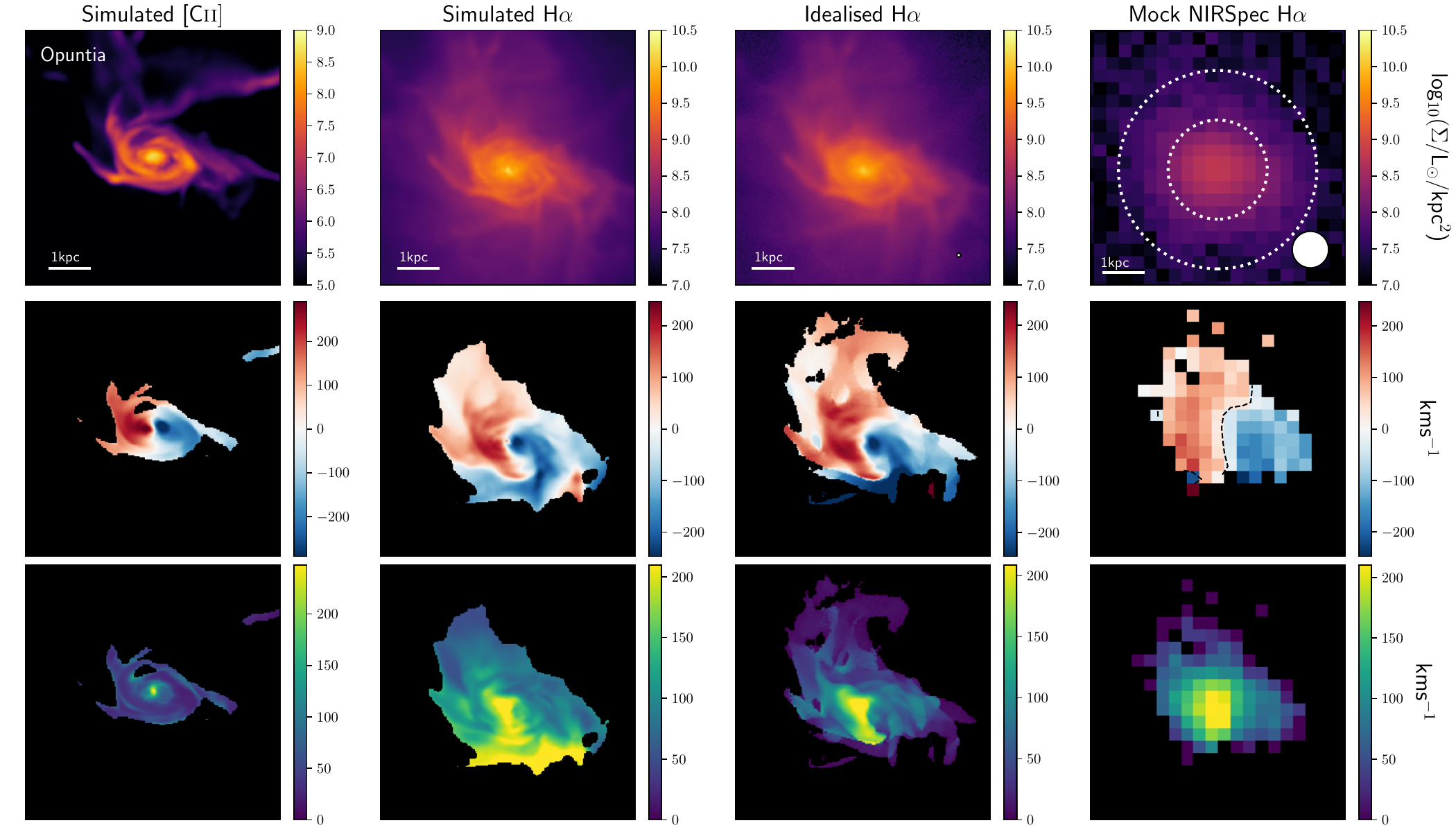}
         \caption{Opuntia}
         \label{fig:Opuntia_maps}
     \end{subfigure}
     \hfill
     \begin{subfigure}{\textwidth}
         \centering
         \includegraphics[width=\textwidth]{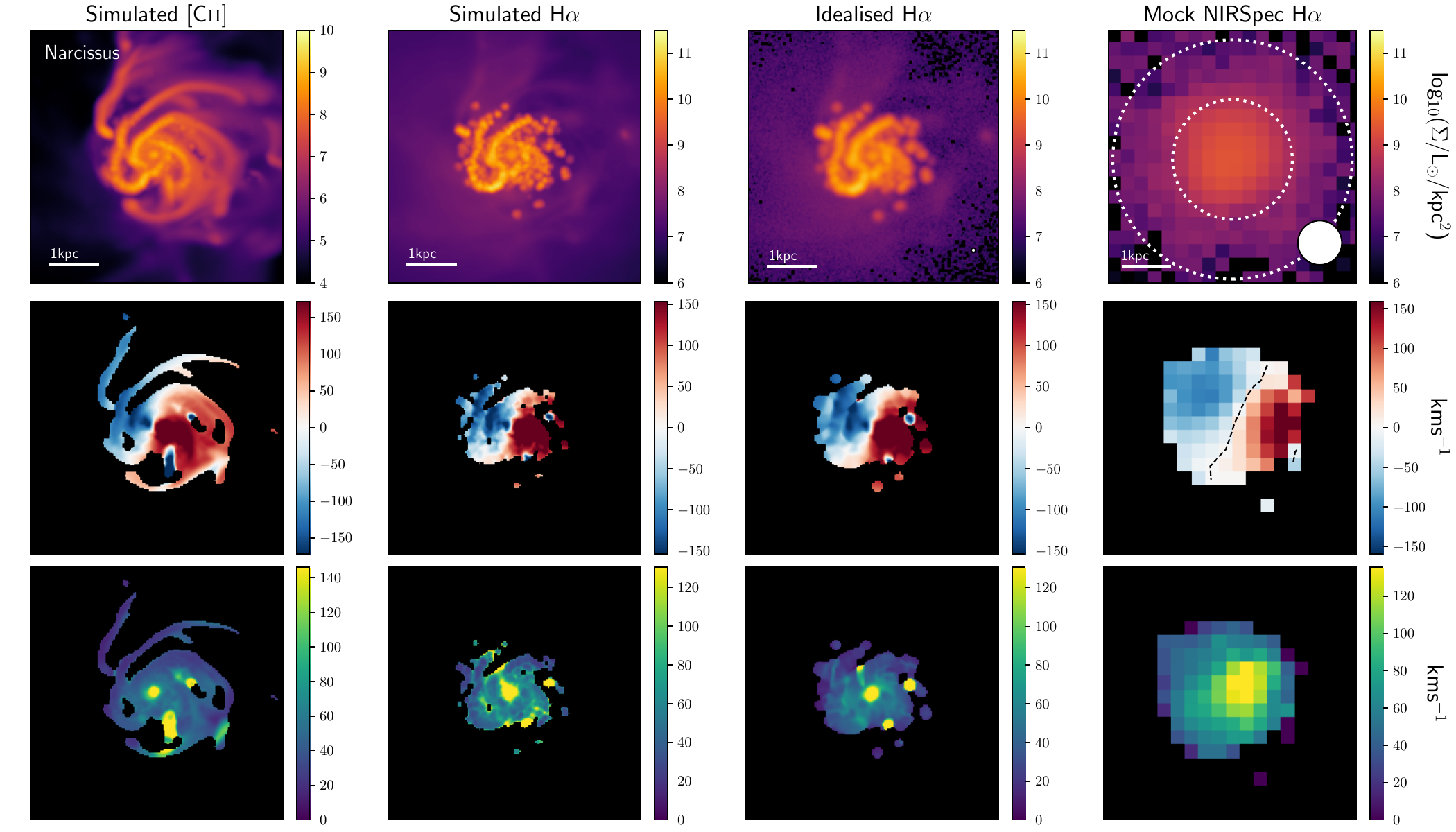}
         \caption{Narcissus}
         \label{fig:Narcissus_maps}
     \end{subfigure}
        \caption{Moment-0, -1, and -2 maps, respectively depicting integrated spectral intensity, line-of-sight velocities, and velocity dispersion for Opuntia (top three rows) and Narcissus (bottom three rows) inclined at 60$\degree$. The [C{\sc ii}] and H$\alpha$ emission maps in the first and second columns are obtained directly from the simulations. See Table \ref{tab:target_properties} for the intrinsic kinematic measurements. The third and fourth columns show the emission line cubes that have been processed to represent an idealised warm gas observation and a realistic mock NIRSpec/IFU observation as described in Section \ref{Creating_Obs}, for which the moment-1 and -2 maps are masked at 25$\sigma$ (in the idealised case) and 5$\sigma$ (in the mock NIRSpec case). The $v=0$ isophote is plotted on the mock NIRSpec moment-1 maps and dotted lines indicate the region representing the `outer extent of the galaxy' in the mock NIRSpec moment-0 maps. PSFs are shown in the moment-0 maps for the derived data.}
        \label{fig:moment_maps}
\end{figure*}

\begin{enumerate}

    \item Spectral binning to achieve a spectral pixel size equivalent to that of the ETC-generated noise cube for each galaxy, which had a value of 32.3 $\kms$ (Narcissus) and 35.8 $\kms$ (Opuntia). Spatial binning to a pixel scale of 0.05"/pixel, representing dithered observations. 

    \item Convolution with a simulated PSF and LSF for NIRSpec/G395H. The PSF has a FWHM of 0.23" (0.20") for Narcissus (Opuntia) and is created with the same dithered pixel size of 0.05", using the python package \texttt{WebbPSF} \citep{Perrin12, Perrin14}, and the LSF is a Gaussian profile with full width at half maximum equivalent to $\lambda$/R, where $\lambda$ is the redshifted wavelength of H$\alpha$ and R is the resolution corresponding to this wavelength, with a value of 3600.8 (3223.9) for Narcissus (Opuntia).

    \item Addition of noise. The ETC-generated noise cube was scaled by a multiplicative factor to reach a SNR of 5 in an aperture with the same area as the PSF at the outer extent of the galaxy. The region referred to here as the `outer extent of the galaxy' is indicated with dotted lines in Figure \ref{fig:moment_maps}, from which it is apparent that the region does not have a uniform flux. To account for this, we use the flux of the entire region to estimate the average signal within a PSF-sized aperture.
    We dropped 10,000 apertures at random positions on random slices of the ETC-generated noise cube, plotted the flux measurements in a histogram and took the standard deviation of the distribution as N. \\
    We fitted the emission line spectrum of the galaxy with a Gaussian profile and extracted the centre ($\mu$) and standard deviation ($\sigma$) of the fitted Gaussian. We used these measurements to create a spectral subcube of the data in the region $\pm 2\sigma$ of the emission line centre. From this subcube we extracted the signal within a PSF-sized aperture at the outer extent of the galaxy, S, and hence obtained the scaling factor by which we multiplied the noise before adding it to the galaxy data. 
    We then estimate the exposure time that would be required to achieve a similar observation of a real target. To do this, we input the properties of our mock observations to the ETC, including the luminosity and effective radius, and calculate the SNR obtained at various exposure times. We find that an SNR of 5 is achievable in an aperture at the outer extent of the galaxy in 7236 seconds for Narcissus (using a detector set-up of 6 groups/integration, 4 integration/exposure, and 4 dithers) and in 57597 seconds for the fainter Opuntia (using a detector set-up of 28 groups/integration, 1 integration/exposure, and 28 total dithers). However, the exposure time required to obtain the same SNR can vary strongly as a result of small differences in target galaxy sizes and emission line widths, the latter being particularly relevant for galaxies with strong velocity gradients or outflow components, as is the case here. Therefore, for our results to translate reliably to real observations, in Section 5.1 we consider the effect on our recovered kinematic measurements of changing the SNR, rather than the exposure time. 
\end{enumerate}

\medskip

\section{Kinematic Analysis} \label{Techniques}

\subsection{Qualitative analysis of the moment maps}

\noindent Moment maps are typically used to obtain evidence on the dynamical state of a galaxy. For instance, the presence of a continuous, smooth gradient in the moment-1 map is used as a criterion in the identification of disks \cite[e.g.][]{Wisnioski15}, but it is not sufficient. At marginal resolution, non-circular bulk motion of gas and interactions between merging systems can mimic certain disk characteristics \citep{Simons19}.

\noindent Figure \ref{fig:moment_maps} shows the moment maps derived from Opuntia and Narcissus. The first and second columns are the [C{\sc ii}] and H$\alpha$ maps obtained directly from the simulated galaxies, the third column is from the idealised observation, and the fourth is from the mock NIRSpec/IFU observation. 

\noindent For Opuntia, the [C{\sc ii}] moment maps trace the dynamically cold central disk, which has a smooth velocity gradient and a centrally peaked velocity dispersion distribution. Opuntia appears far more extended in H$\alpha$, where we still see the central disk structure but also the diffuse gas surrounding it. 
In the mock NIRSpec/IFU observation, the moment-1 map shows the smooth gradient indicative of a disk. However, the velocity gradient is not only tracing the disk structure itself. The comparison with the intrinsic simulated H$\alpha$ maps (second column) clearly shows that the velocity gradient in this case is tracing both the disk and the surrounding gas. Hence, at the angular resolution of the JWST observations, the diffuse component dominates and mimics the emission originating from the outer extent of the galaxy. The deviation of the $v=0$ isophote from a straight line in the centre of the moment-1 map provides an indication of the presence of non-circular motion \citep[e.g.][]{Arribas24, Ubler24}. However, this method can be misleading in certain cases, as the features in the velocity field can be smoothed out when the angular resolution is low \citep{Rizzo22}, and the shape of the isovelocity contour can vary strongly and arbitrarily depending on the mask that is applied to produce the velocity field (Appendix \ref{appendix-ganifs}).\\
Traced by [C{\sc ii}], we see that the central disk of Narcissus is encircled by extended filiamentary structures. From their morphology in the moment-0 map alone, these could be interpreted as inflowing or outflowing gaseous streams, but the moment-1 map appears to contradict this scenario as the filiaments reproduce the velocity gradient of the inner disk. We therefore consider them to be gravitationally bound to the disk, and perhaps best described as \enquote{cold gas spiral arms}.
For Narcissus, the structure traced by cold and warm gas appears similar; according to both, Narcissus has a strong axial asymmetry in the [C{\sc ii}] and H$\alpha$ distributions. Its irregular structure comprises multiple off-centre clumps clearly visible in the moment-2 maps, which could represent satellites or localised clumps of star formation. While there is a velocity dispersion peak coincident with the galactic centre, the flux is not centrally concentrated and is instead weighted to the south of the galaxy. In the mock NIRSpec/IFU moment-1 map we again see a smooth, disky velocity gradient, though here it is indeed tracing disk structure. The high-velocity dispersion clumps are no longer resolved, and appear merged into a single off-centred region. 

\subsection{Kinematic Modelling with \texorpdfstring{\textsuperscript{3D}\textsc{Barolo}}{3D BAROLO}}
\label{barolo}
\noindent Kinematic fitting to the simulated IFU observations is performed using the software \textsuperscript{3D}\textsc{Barolo} \citep{Barolo15}, which fits 3D tilted ring models \citep{Rogstad74} to emission line datacubes. \textsuperscript{3D}\textsc{Barolo} creates mock realizations of rotating disk models defined by kinematic and geometrical parameters, notably including rotational velocity, velocity dispersion, inclination angle, position angle, and the coordinates of the disk centre. 

\noindent Before comparing the data with the model, \textsuperscript{3D}\textsc{Barolo} convolves the model with a Gaussian component which approximates the observational PSF to account for the impact of beam smearing. \textsuperscript{3D}\textsc{Barolo} outputs model cubes convolved with the beam, as well as moment maps and major and minor axis position-velocity (PV) diagrams, which are 2D velocity profiles extracted along the major and minor kinematic axes of a source.
To perform fitting with \textsuperscript{3D}\textsc{Barolo}, we provide estimated values for certain parameters motivated by measurements and assumptions of the underlying physics, as described in the following. 

\smallskip
\textit{Kinematic Parameters}. 
The rotation velocity and velocity dispersion are allowed to vary between 10-300 $\kms$, with initial parameter estimates of 250 $\kms$ and 50 $\kms$ respectively, informed by measurements for rotating disk candidates at similar redshifts \citep{Smit18, Posses23, Rowland24}. 

\smallskip
\textit{Geometrical Parameters}.  
Since the goal of this project is to constrain the impact of data quality on the feasibility of deriving dynamical properties, it is a \enquote{proof of concept} study and we therefore did not consider it necessary to fit the inclination angle from the data as if we had no \textit{a priori} knowledge of the galaxy properties.

Position angle is left to be fitted. The \texttt{Z0} parameter, controlling the scale height of the disk, is fixed at 0.001" for each observation as we assume a thin disk. We experiment with altering the assumed disk thickness and confirm that this has very little effect on the fitted model, as the rotating disk features are consistent between thin and thick discs and the uncertainties introduced by the thin disk assumption is not significant compared to the errors on the velocity measurements. Assumed disk thickness mainly affects the inclination angle \citep{R-O23} which for our purposes is fixed. 

\smallskip
\textit{Resolution Elements}. The radial separation (\texttt{RADSEP}) was set so as to obtain the maximum possible number of independent measurements without oversampling. The value of \texttt{NRADII} is then the number of independent tilted rings that could be placed along the extent of the galaxy at a separation of \texttt{RADSEP} $\approx$ FWHM. For the realistic mock NIRSpec observations, \texttt{NRADII} is 2. 

\smallskip
\noindent In Appendix \ref{channelmaps} we present representative channel maps of each datacube along with the best fitting model from \textsuperscript{3D}\textsc{Barolo}.

\section{Results}
\label{results_section}

\noindent In this section, we introduce evidence for the presence of disks in our sample (Section \ref{diskid}), and examine the impact of observational effects on the recovered kinematics and related measurements: turbulence and rotational support (Section \ref{measurementbias}).
We use our mock observations to comment on the validity of current techniques used for outflow identification (Section \ref{outflowsection}), and furthermore we test the impact of the biases in our kinematic measurements on the recovered dynamical masses (Section \ref{dynamicalmass}).

\subsection{Disk Identification}
\label{diskid}

\begin{figure}
    \centering
    \includegraphics[width=\columnwidth]{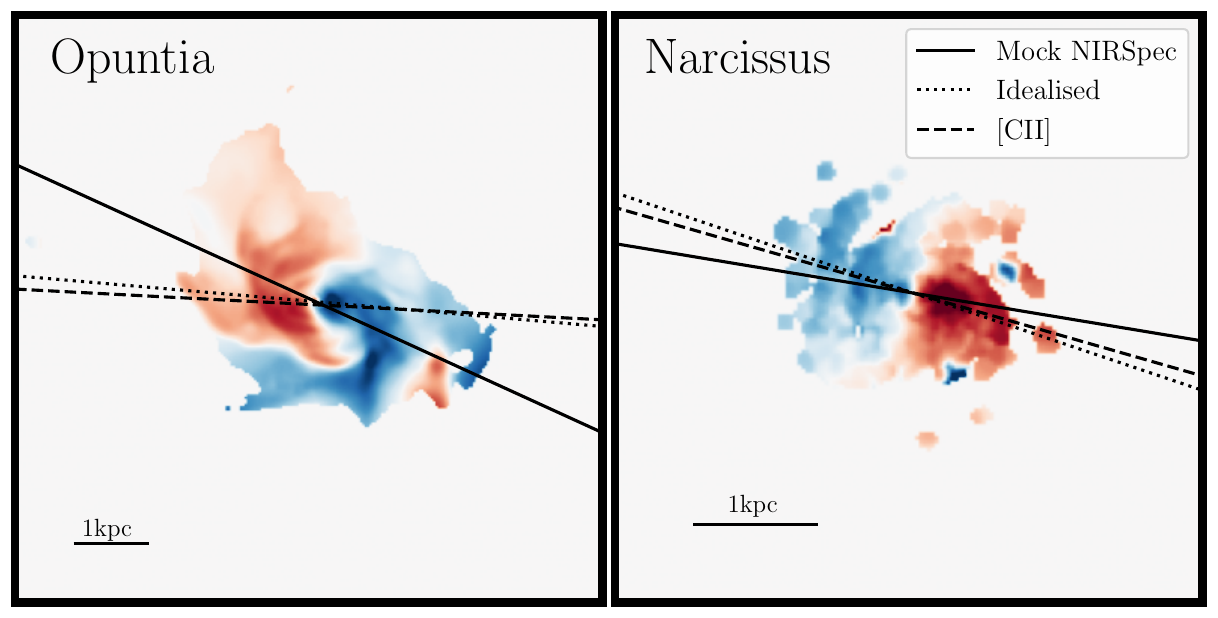}
    \caption{Kinematic axes obtained from \textsuperscript{3D}\textsc{Barolo} fitting (section \ref{barolo}), measured from the realistic mock NIRSpec observations, along with idealised H$\alpha$ and [C{\sc ii}] observations, overlaid on the intrinsic H$\alpha$ moment-1 maps.}
    \label{fig:m0pas}
\end{figure}

\begin{figure*}
     \centering
     \begin{subfigure}{0.48\textwidth} 
         \centering
         \includegraphics[width=\textwidth]{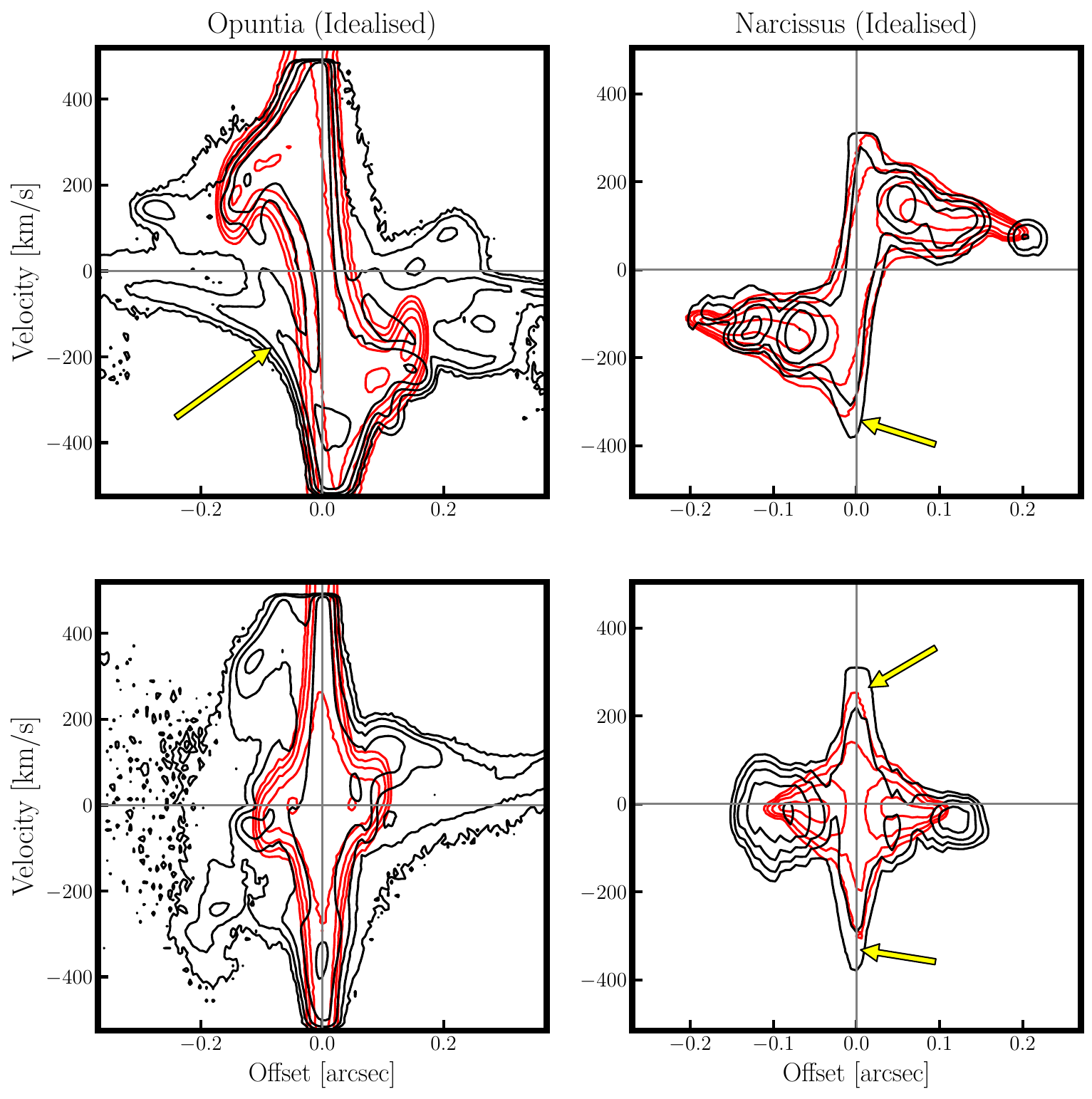}
          \label{fig:ideal_pv}
     \end{subfigure}
     \hspace{0.02\textwidth} 
     \begin{subfigure}{0.48\textwidth} 
         \centering
         \includegraphics[width=\textwidth]{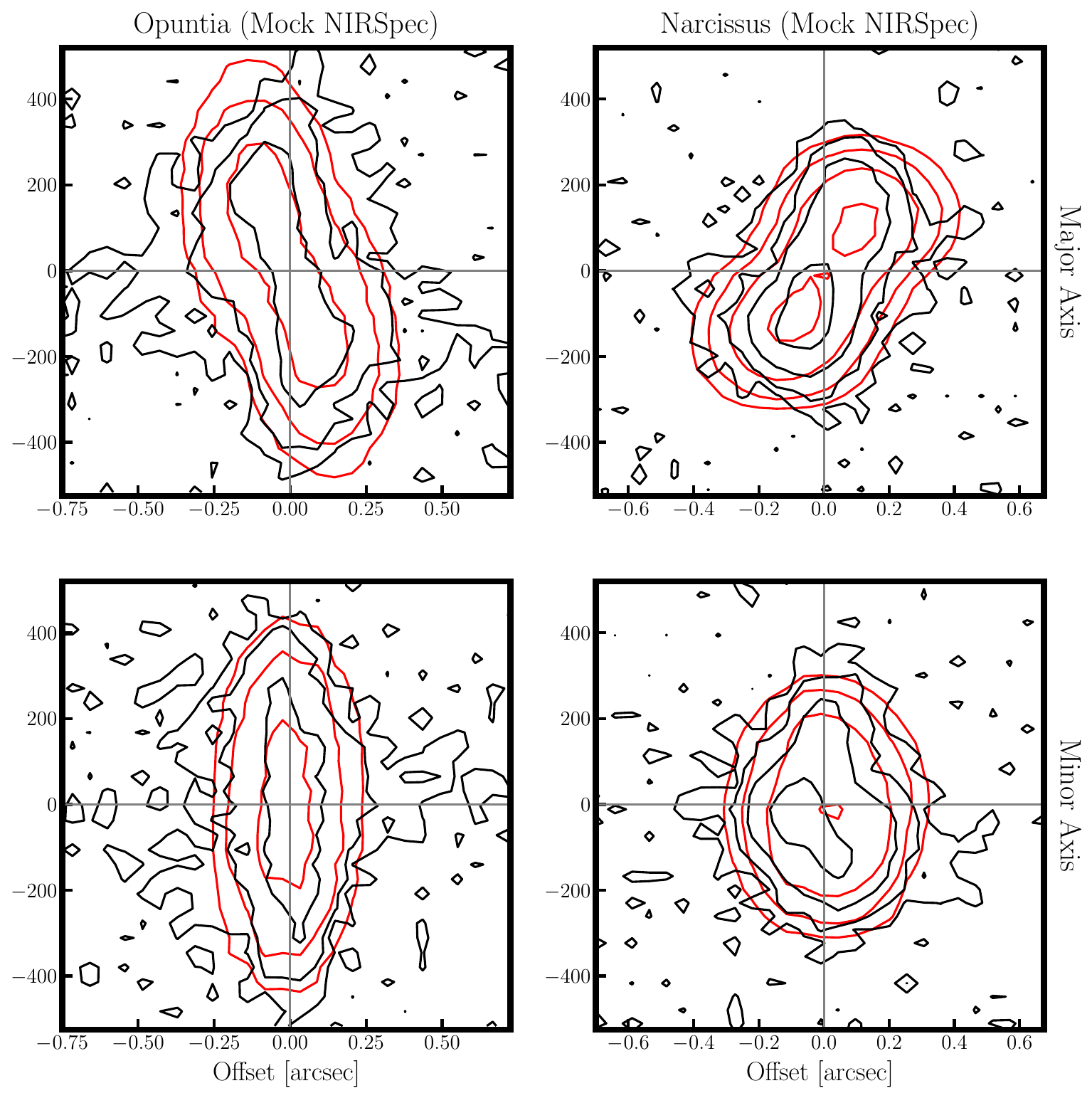}
         \label{fig:mock_pv}
     \end{subfigure}
        \caption{Position-Velocity diagrams for the idealised observations (left) and mock NIRSpec observations (right), where the major axis diagrams are extracted along the axes shown in Figure \ref{fig:m0pas}. Contour levels are at $3^n\sigma$, where n=[1, 2, 3, 4, 5] and the $\sigma$ value used to define the contours is the RMS value of noise-dominated regions in the diagrams. Black contours trace the data, and red contours represent the disk model. Arrows indicate regions of the PV diagrams that deviate from the expected profile for a disk.}
        \label{fig:PVs}
\end{figure*}

\noindent We use the position angle fitted by \textsuperscript{3D}\textsc{Barolo} to the mock NIRSpec observations and idealised H$\alpha$ and [C{\sc ii}] observations to define the kinematic axis as measured by each tracer.
Each measured axis is plotted over the corresponding galaxy in Figure \ref{fig:m0pas}. We extract the major PV diagrams along these kinematic axes, and the minor PV diagrams along the orthogonal axes. In Figure \ref{fig:PVs}, we show the contours of the data (black) and the disk model (red). PV diagrams provide evidence for determining the kinematic properties of galaxies, with a rotating disk showing a characteristic signature in its PV diagrams \citep{Fraternali02, deBlok08, Neeleman20}.
The minor axis diagram of a rotating disk is symmetric about the axes defining the systemic velocity and the centre, and the major axis diagram displays an S-shape \citep{Rizzo22}. For a rotating disk with no outflows, inflows, or other sources of non-virial motions, there should be no emission outside of the S-profile, and therefore the quadrants unoccupied by the S-profile are referred to as \enquote{forbidden regions}.

\noindent For Opuntia, it is possible to discern a disk-like S-shaped distribution in the major-axis PV diagram of both the idealised and realistic mock NIRSpec cases, though it is significantly less distinct in the mock NIRSpec diagram due to the lower resolution. In the idealised case, it is clear that there are non-circular motions present as there is emission not reproduced by the rotating disk model visible in the so-called forbidden regions (e.g. the emission region in the negative quadrant of the major axis PV diagram, indicated with an arrow). In the mock NIRSpec case, the rotating disk model is able to well reproduce the overall emission in the bright inner region, but the agreement between the data and disk model worsens for the faint emission (the outermost contour line), in both the major and minor axis PV diagrams. 

\noindent In Narcissus, it is evident that the major axis PV diagrams are heavily luminosity-weighted to one side, inconsistent with what would be expected for an axisymmetric disk. In the idealised case there is a small peak that could potentially be associated with an outflow, extending below the disk profile S-shape in the inner region, at approximately $-350$ $\kms$ (indicated on Figure \ref{fig:PVs} with an arrow), and a similar feature is also visible in the minor axis diagram (again indicated by an arrow).

\noindent The complementarity of data and model in the mock NIRSpec PV diagrams is poorer still for Narcissus than for Opuntia. Despite Narcissus being less influenced by non-circular motions, the asymmetry of the H$\alpha$ distribution and the presence of bright clumps lead to a significant discrepancy between data and disk model along both the major and minor axis.  From the mock NIRSpec observations, both galaxies would be identified as disks, Opuntia with $v/\sigma = 2.1$, and Narcissus with $v/\sigma = 3$ (Table \ref{tab:velocity_vals}). However, neither would be classified as cold disks, despite Narcissus being a dynamically cold disk, and Opuntia being a disk with strong rotational support (i.e. v/$\sigma$ > 5) in simulated H$\alpha$.

\begin{figure}
	\includegraphics[width=0.5\textwidth]{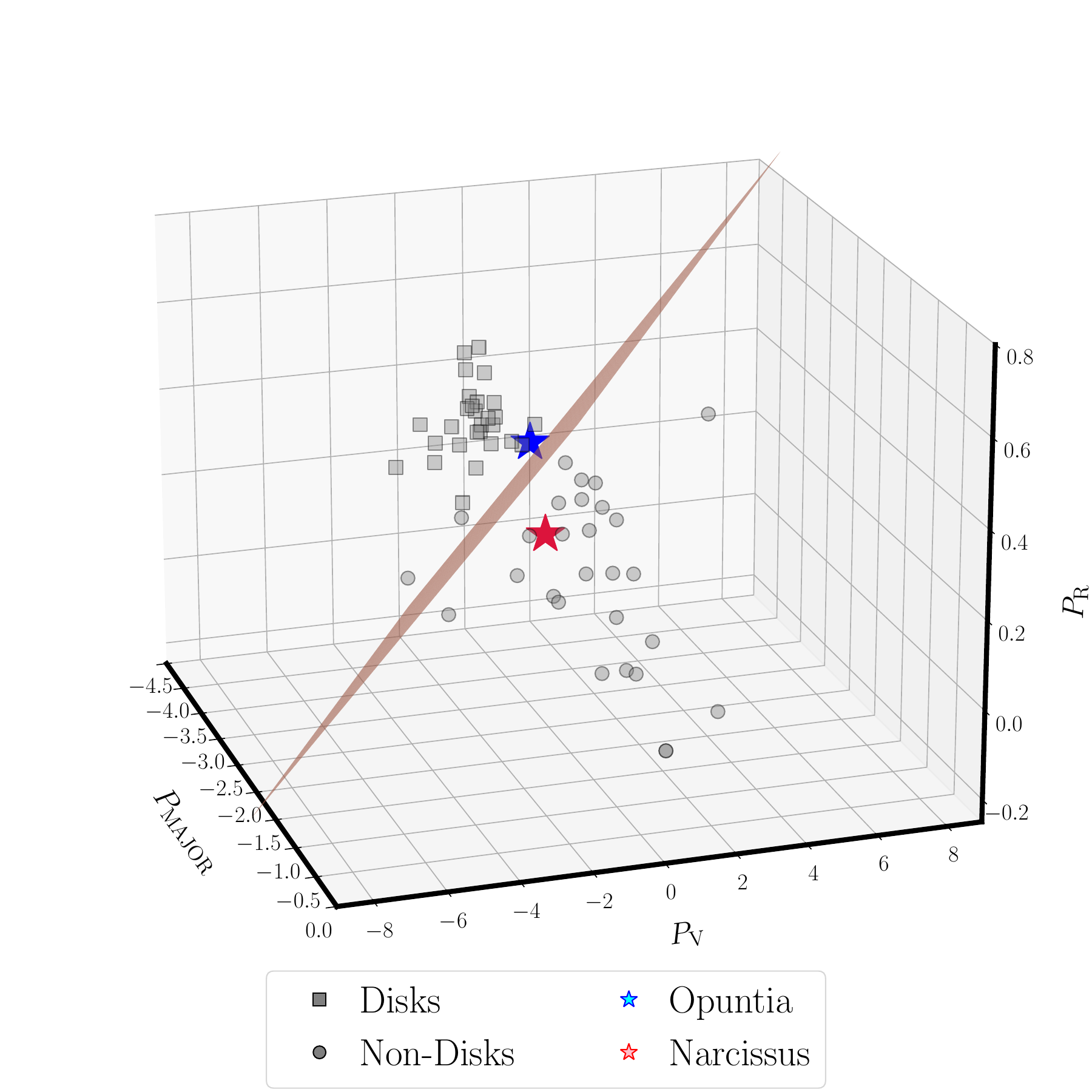}
    \caption{Narcissus and Opuntia are here plotted in the PVSplit parameter space alongside simulated disks and non-disks \citep{Rizzo22}. Narcissus occupies the non-disk region of parameter space, while Opuntia is close to the plan of division between the disk and non-disk populations \citep{R-O23}. Uncertainties in the PVSplit fitting, propagated from uncertainties in e.g. the fitted centre and position angle from \textsuperscript{3D}\textsc{Barolo}, tend to act towards the non-disk region.}
    \label{fig:pv3d}
\end{figure}

\noindent It is evident that the presence of non-circular gas motion in Opuntia, and of irregular disk structure in Narcissus, affects the shapes of their PV diagrams. 
We therefore examine the ensuing consequences for their disk classification according to the PVSplit method, a dynamical classification technique introduced in \cite{Rizzo22}.
PVSplit analysis is based on three empirical parameters that quantify the symmetric and morphological properties of the major and minor PV diagrams. These are P$_\text{major}$, a measure of the asymmetry of the major axis PV diagram with respect to the axis defining systemic velocity, P$_\text{V}$, which quantifies the distribution of emission peaks along the velocity axis, and P$_\text{R}$ which is similarly defined along the radial axis. The accuracy of PVSplit in separating mergers and disks is demonstrated in \cite{Rizzo22} for mock ALMA observations at SNR $\geq 10$ where at least 3 independent resolution elements can be laid along the galactic major axis.
Figure \ref{fig:pv3d} shows where Narcissus and Opuntia fall on the PVSplit parameter space in relation to the plane of best division between disk and non-disk systems, as defined by \cite{R-O23} using the support-vector machine method to maximize the distance separating the disk and non-disk galaxy samples pre-classified by \cite{Rizzo22}, which are also plotted on the PVSplit diagram. Opuntia is classified as a disk, while Narcissus is classified as non-disk. 

\noindent In summary, the comparison between the data and disk model reveals that the two galaxies might not be identified as regular rotating disks from realistic NIRSpec/IFU observations. This result indicates that despite the limited angular and spectral resolution, the mock observational data is not accurately reproduced by an axisymmetric disk model, and rather we see that the presence of non-circular motions in Opuntia and of asymmetries in Narcissus are identifiable through discrepancies with respect to the disk model. \\

\noindent We tested the effect on our recovered rotational velocity and velocity dispersion of changing the SNR from $\sim 5$ to SNR$\sim 3$ and $\sim 10$, finding that the velocity measurements are robust for both galaxies across the entire S/N range. The kinematic measurements recovered at SNR$\sim 3$ and SNR$\sim 10$ are presented in a table in Appendix \ref{appendix_resolution}.

\subsection{Quantifying Biases in Measured Turbulence and Rotational Support}
\label{measurementbias}

\begin{figure}
	\includegraphics[width=\columnwidth]{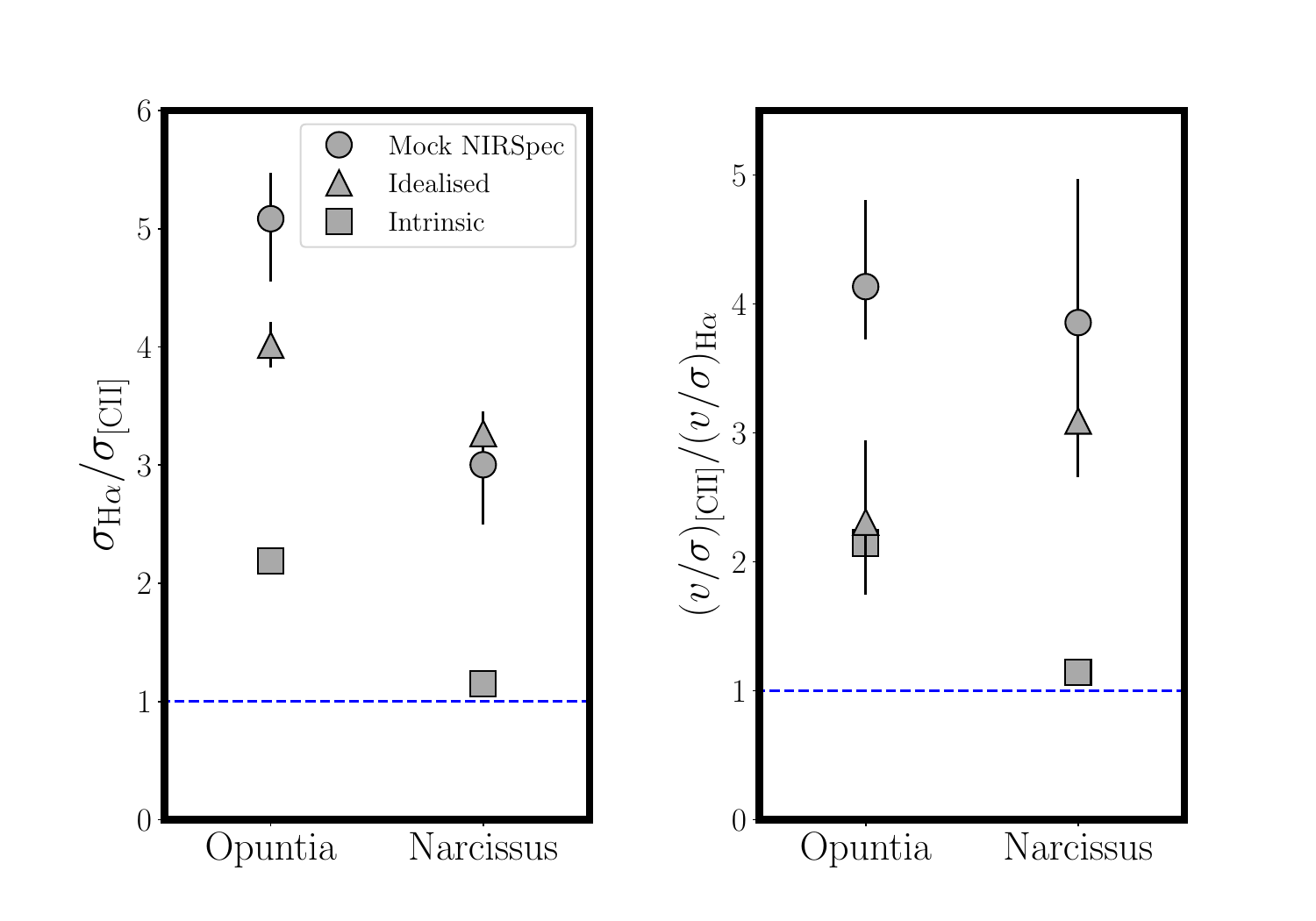}
    \caption{v/$\sigma$ and $\sigma$ values from the intrinsic, idealised and mock NIRSpec H$\alpha$ measurements as a function of the intrinsic [C{\sc ii}]} measurement.
    \label{fig:vsigma}
\end{figure}

\noindent From the \textsuperscript{3D}\textsc{Barolo} fitting described in Sections \ref{barolo} and \ref{diskid}, we extracted the rotation velocity and velocity dispersion profiles. We thus obtained the radial average values of $\sigma$ and $v/\sigma$, which we computed as the ratio between the maximum rotation velocity and the average velocity dispersion across the individually fitted rings. Considering each of these in comparison to the intrinsic [C{\sc ii}] measurement gives insight into the extent to which the presence of non-circular motions under the conditions of lower spectral and angular resolution amplifies the disparity between the turbulence and rotational support measured from warm and cold gas.  Figure \ref{fig:vsigma} shows the comparison between these values and their analogues calculated directly from the simulations in both [C{\sc ii}] and H$\alpha$, as described in Section \ref{targets}. The values are tabulated for both the idealised and realistic mock NIRSpec data in Table \ref{tab:velocity_vals}. In the following, we describe the comparison of velocity measurements extracted from our idealised and mock observational data with those from the intrinsic simulated H$\alpha$ and [C{\sc ii}]. This comparison is performed to probe the impact of the gas tracer, without introducing complexity through observational effects acting on the reference values themselves.

\begin{table}
	\centering
	\caption{The maximum values of rotational velocity and the average values of the velocity dispersion for each galaxy, as ascertained with \ce{^{3D} B{\sc arolo}, alongside the intrinsic measurements}. }
	\label{tab:velocity_vals}
        \renewcommand{\arraystretch}{1.5}
	\begin{tabular}{| p{3.1cm} | p{1.3cm} p{1.3cm} p{1cm}|}
	\hline
         & v$_\text{rot, max}$ [$\kms$] & $ \langle \sigma \rangle$ [$\kms$] & v/$\sigma$ \\
        \hline
	Opuntia (Idealised) & $410_{-85}^{+89}$ & $108_{-5}^{+5}$ & $3.8_{0.8}^{0.8}$\\
        Opuntia (Mock NIRSpec) & $290_{-28}^{+25}$  & $136_{-14}^{+10}$ & $2.1_{-0.3}^{+0.2}$\\
        Opuntia (Intrinsic H$\alpha$) & 240.3 & 58.6 & 4.1\\
        Opuntia (Intrinsic [C{\sc ii}]) & 235.8 & 26.8 & 8.8\\
        Narcissus (Idealised) & $261_{-27}^{+23}$  & $68_{-3}^{+3}$ &  $3.8_{-0.5}^{+0.5}$\\
        Narcissus (Mock NIRSpec) &  $193_{-17}^{+14}$  & $62_{-11}^{+10}$ & $3_{-0.6}^{+0.9}$\\
        Narcissus (Intrinsic H$\alpha$) & 248.6 & 23.9 & 10.4\\
        Narcissus (Intrinsic [C{\sc ii}]) & 247.5 & 20.8  & 11.9 \\
	\hline
  
	\end{tabular}
\end{table}

\noindent In the case of Narcissus, less dominated by non-circular motions, the H$\alpha$ $\sigma$ ($v/\sigma$) values from the realistic mock NIRSpec and idealised cases are consistent with each other, but $\sim 3 \times$ higher ($\sim 3 \times$ lower) than the intrinsic H$\alpha$ measurement, which is close to the intrinsic [C{\sc ii}] measurement. We thus interpret the difference with respect to the intrinsic value as due to the assumption of an axisymmetric disk that we adopted to derive the kinematics, since it is apparent in both the high and low resolution observations. 

\noindent For Opuntia, in the idealised case, the measured $\sigma_{\text{H}\alpha}$ (v/$\sigma_{\text{H}\alpha}$), is $1.8\times$ higher ($1.1 \times$ lower) than the intrinsic $\sigma_{\text{H}\alpha}$ and $4.0\times$ higher ($2.3\times$ lower) than $\sigma_\text{[CII]}$ (v/$\sigma_\text{[CII]}$). For the mock NIRSpec observations, the measured $\sigma_{\text{H}\alpha}$ (v/$\sigma_{\text{H}\alpha}$) is $2.3\times$ higher ($2.0\times$ lower) than the intrinsic $\sigma_{\text{H}\alpha}$ and $5.1\times$ higher ($4.2\times$ lower) than $\sigma_\text{[CII]}$ (v/$\sigma_\text{[CII]}$). These comparisons illustrate the extent to which the non-circular gas component artificially increases the turbulence and decreases the rotational support as measured by hot gas relative to cold gas intrinsically. For both the idealised and the mock NIRSpec observations, a rotating disk model is not able to well reproduce the data. In the idealised case, this is very clear, whereas for the mock NIRSpec case, we ascertain the disagreement between data and model from  indications based on the fitting in the lower SNR outer regions of the galaxies, and the PVSplit results. Given the limited constraints available for the mock NIRSpec observations, making improvements to the modelling by adding, for example, non-circular motions, will not improve fitting due to the high number of parameters to be fitted relative to the available observational constraints. The higher turbulence and lower rotational support measured from the mock NIRSpec observations compared to the idealised observations motivates the conclusion that the lower resolution of the mock NIRSpec observations, as well as a residual beam smearing effect \citep[e.g.][]{Bosma78, Begeman87, Barolo15, Zhou17} and contamination by non-circular motions, leads to artificially inflated values of the velocity dispersion. 
Under the assumption of a rotating disk, \textsuperscript{3D}\textsc{Barolo} tries to reproduce the diffuse gas in non-circular motions by increasing the velocity dispersion, and therefore the measured $\sigma$ values are biased upwards.

\subsection{Recovering Outflow Properties from Observations}
\label{outflowsection}
\noindent Opuntia has a significantly more turbulent H$\alpha$ component, a phenomenon which was connected to the presence of outflows in \cite{Kohandel24}, and further developed in \cite{Kohandel25}. Due to the complex geometry of outflow host candidates, it is challenging to positively attribute non-circular motion of gas observed in galaxies to the presence of outflows \citep[for a review of outflow detection techniques see][]{Veilleux20}. However, in this section we treat the presence of outflows in Opuntia as confirmed, a result which will be analysed in forthcoming work on outflows in simulations. 
\noindent To understand the properties of the gas surrounding the disk in Opuntia, we separate the disk and non-disk emission in the idealised case using the disk model fitted by \textsuperscript{3D}\textsc{Barolo}. From the disk model, we create a disk mask consisting of all pixels included in the disk model on a channel-by-channel basis, and we invert this to create a non-disk mask. 
We apply these masks in turn to the Opuntia idealised observation, yielding two separate datacubes consisting of only disk emission and only non-disk emission. Maps of these separate emission components are presented in Appendix \ref{separated_emission_components}.
In Figure \ref{fig:outflow} the integrated spectrum of the idealised observation is plotted in black, and its disk and non-disk components in purple and pink respectively. From this, we additionally show the PV diagrams for each component extracted along the same axis as was used for the PV diagram of the unmasked idealised observation. We see that the integrated spectrum of the entire idealised observation has a Gaussian profile. The non-disk emission component shows a similar Gaussian profile with a lower amplitude and full width at half maximum, and the integrated spectrum of the disk component exhibits a characteristic double-horned peak profile. 

In literature, a widely used technique to identify outflows is fitting a double Gaussian profile to the emission line spectrum of a galaxy, comprising one narrow component representing virial motion, and one broad component associated with outflows \citep[e.g.][out of many examples.]{Carniani15, Maiolino17, HC21} In Appendix \ref{doubleGaussian} we attempt to fit a two-component Gaussian composite line profile to the integrated spectrum of the mock NIRSpec observations, and find that the fitted Gaussian peaks do not conform to the broad and narrow shapes that would generally be considered indicative of the presence of outflows. 
\cite{Concas22} criticise the oversimplifying assumptions inherent in the double Gaussian fitting outflow detection method, which does not account for observational effects which may artificially give rise to a non-Gaussian emission line profile that could include broad wings. The authors caution that this method may result in the identification of outflows where none are present. 

\begin{figure}
	\includegraphics[width=0.5\textwidth]{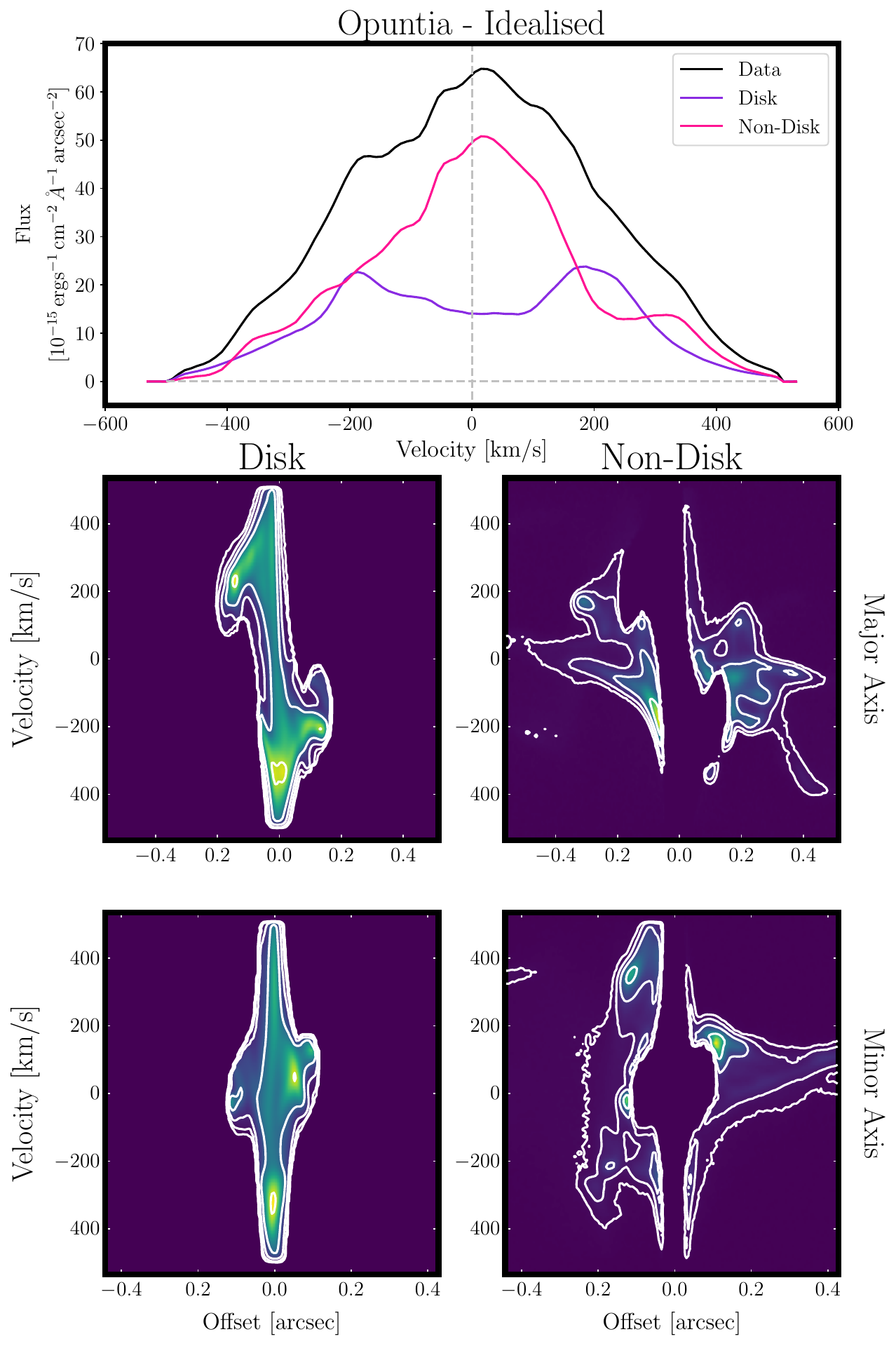}
    \caption{\textit{Top Figure:} The spectrum of the Opuntia idealised observation is plotted as a black line. The total flux is split into a disk and non-disk component. The spectra of these components are plotted as purple and pink lines respectively. \\
    \textit{Second Row:} Major axis PV diagrams for the disk and non-disk components, extracted along the idealised observation axis identified in Figure \ref{fig:m0pas}. Contour levels are at $3^n\sigma$, beginning at $n=1$. \\
    \textit{Third Row:} As above for the minor axis PV diagrams, contour levels at $3^n\sigma$.}
    \label{fig:outflow}
\end{figure}

\noindent Our results as presented in Figure \ref{fig:outflow} add an additional caveat; this method may also fail to identify outflows where they are in fact present. The integrated spectrum of the idealised observation does not display the profile of a double Gaussian with a narrow and broad component despite hosting diffuse emission, likely due to outflows, which dominates the emission from the galaxy. This component is only revealed by the 3-dimensional analysis made possible by the IFU, and thus our result indicates a challenge for confirming outflows from integrated or slit spectroscopy.

\subsubsection{Considering Disk and Non-Disk Emission at Low Resolution}

\noindent Applying the method described above to separate the idealised observation of Opuntia into a disk and non-disk component and then comparing the summed emission in each gives a ratio of the disk to non-disk emission of 1:1.9. If we apply the same method to separate the emission components in the mock NIRSpec observation of Opuntia, we obtain a disk to non-disk emission ratio of 1:0.3, signifying the extent to which low resolution causes non-disk emission to appear part of the disk. Appendix \ref{channelmaps} provides the fitted channel maps and residuals for both the idealised and mock NIRSpec observations. Residuals may be thought of as identifying the non-disk emission; however, in the mock NIRSpec case and for real observations with low SNR, this would be an over-interpretation of the data, as with decreasing SNR it becomes more challenging to distinguish real emission from noise. 

\subsection{Implications for Dynamical Mass}
\label{dynamicalmass}

\noindent The observational and modelling biases that affect kinematic measurements may consequently impact the estimates of the total mass budget within a galaxy. The rotational velocity is, in fact, a tracer of the galactic gravitational potential ($\Phi$), which additionally depends on the pressure support provided by random motion within the gas (i.e. $\sigma$) \citep{BinneyTremaine87, Cimatti20}.

\begin{equation}
 \frac{1}{\rho} \frac{\partial \rho \sigma^2}{\partial R} = -\frac{\partial \Phi}{\partial R} + \frac{v^2_\text{rot}}{R},
\end{equation}
where $R$ is the radius and $\rho$ is the gas volumetric density. This equation can be rewritten as
$v^2_\text{circ} = v^2_\text{rot} + v^2_A$, where $v_\text{circ}$ is the circular velocity, defined by  $v^2_\text{circ} = R\partial \Phi/ \partial R$ and $v_\text{A}$ is the asymmetric drift correction
 \begin{equation}
 \label{eq:va}
 v^2_A = -R\sigma^2 \frac{\partial \ln(\rho \sigma^2)}{\partial R}. 
 \end{equation}

When the galaxy is dynamically cold ($V/\sigma \gtrsim 10$), the asymmetric drift term is negligible and $v_\text{rot} \approx v_\text{circ}$. In this case, to estimate the order of magnitude of the total mass enclosed up to outermost observable radius $R_\text{out}$, we assume spherical symmetry and define the dynamical mass as 

\begin{equation}
    M_\text{dyn} = \frac{R_\text{out} v^2_\text{circ}}{G} \approx \frac{R_\text{out} v^2_\text{rot}}{G}\, .
\end{equation}
For galaxies with low rotation support ($V/\sigma < 10$), the asymmetric drift correction should be applied to obtain an unbiased estimate of the total gravitational potential. For a galaxy disc where the gas thickness does not depend on radius, and for an exponential surface brightness profile $\Sigma = e^{-R/R_{gas}}$, the asymmetric drift term can be written as 
\begin{equation}
\label{eqn_asymdrift}
v^2_A = -2R\sigma^2 \frac{\partial ln(\sigma)}{\partial R} + \frac{R\sigma^2}{R_\text{d, gas}},
\end{equation}
where $R_\text{d, gas}$ is the scale radius of the disc \citep{Roman-Oliveira24}.

\noindent Both Opuntia and Narcissus have low rotation support estimated from the mock NIRSpec observations, so the asymmetric drift term will be non-negligible. Despite this, we calculate dynamical mass using both the asymmetric drift-corrected and uncorrected formalisms to quantify the accuracy of the dynamical mass measurements in the case in which the measured $\sigma$ values are overestimated. The process by which dynamical masses are calculated is expounded on in Appendix \ref{appendix_mdyn}. In Table \ref{tab:masstable}, we show the total masses within 1kpc measured directly from the simulations, including stars, gas and dark matter, which are $1.8  \times 10^{10} \text{M}_{\odot}$ and $2.1 \times 10^{10} \text{M}_{\odot}$ for Opuntia and Narcissus respectively. 

\renewcommand{\arraystretch}{1.6}
\begin{table}
    \centering
    \caption{Mass values within 1kpc measured from the simulations, the mock NIRSpec observations, and the idealised observations. }
    \label{tab:masstable}
    \begin{threeparttable}
    \begin{tabular}{|p{3.5cm}|p{1.5cm}|p{1.5cm}|}
        \hline
		   & Opuntia & Narcissus \\
                \hline
             Intrinsic $M_\star$ [$10^{10}$ M$_{\odot}$] & $1.1 $ & $1.0$ \\
             Intrinsic M$_{\textrm{gas}}$ [$10^{10}$ M$_{\odot}$] & $0.24$ & $0.41 $ \\
             Intrinsic M$_\textrm{DM}$ [$10^{10}$ M$_{\odot}$] & $0.49 $ & $0.72 $ \\
             Intrinsic M$_\textrm{total}$ [$10^{10}$ M$_{\odot}$] & $1.8$ & $2.1$ \\
             \hline
             Idealised M$_\text{dyn}$ [$10^{10}$ M$_{\odot}$] \tnote{a} &  $ 1.4^{+0.3}_{-0.3} $ &   $0.9_{-0.4}^{+0.3}$ \\
             Idealised M$_\text{dyn}$ [$10^{10}$ M$_{\odot}$] \tnote{b} & $ 1.7^{+0.3}_{-0.3} $  &   $0.9_{-0.4}^{+0.3}$\\
             \hline
             Mock NIRSpec M$_\text{dyn}$ [$10^{10}$ M$_{\odot}$] \tnote{a} & $1.3^{+0.5}_{-0.5} $   & $0.8_{-0.3}^{+0.3}$\\
             Mock NIRSpec M$_\text{dyn}$ [$10^{10}$ M$_{\odot}$] \tnote{b} & $2.3^{+0.6}_{-0.6} $  &   $1.0_{-0.3}^{+0.3}$\\
             \hline
             Idealised integrated spectra M$_\text{dyn}$ [$10^{10}$ M$_{\odot}$] & $3.22^{+0.3}_{-0.3}$ & $2.5^{+1.5}_{-1.5}$ \\
             \hline
             Mock NIRSpec integrated spectra M$_\text{dyn}$ [$10^{10}$ M$_{\odot}$] & $4.27^{+0.6}_{-0.6} $ & $1.6^{+0.6}_{-0.6}$\\

        \hline
    \end{tabular}
    \begin{tablenotes}
        \item[a] Approximating rotational velocity and circular velocity as equivalent: $v_\text{rot} \sim v_\text{c}$. 
        \item[b] Applying the correction for asymmetric drift (equation \ref{eqn_asymdrift}). 
    \end{tablenotes}
    \end{threeparttable}
\end{table}

\noindent For the idealised observations, the dynamical masses obtained without the asymmetric drift correction are a factor of 1.3$\times$ (Opuntia) and 2.3$\times$ (Narcissus) lower than the intrinsic values derived directly from the simulations (see Table \ref{tab:masstable}). When the asymmetric drift correction is included the dynamical mass estimate becomes 1.1$\times$ below the intrinsic value for Opuntia, within 1$\sigma$ uncertainty, and is unchanged at 2.3$\times$ below for Narcissus, within 4$\sigma$ uncertainty. 

\noindent For the mock NIRSpec observations, the dynamical masses obtained without the asymmetric drift correction are a factor of 1.4$\times$ (Opuntia) and 2.6$\times$ (Narcissus) lower than the intrinsic values derived directly from the simulations (see Table \ref{tab:masstable}). When the asymmetric drift correction is included the dynamical mass estimates are 1.3$\times$ above and 2.1$\times$ below the intrinsic values, within 1 and 4$\sigma$ uncertainties, for Opuntia and Narcissus respectively. 

\noindent Therefore, we see that including the asymmetric drift correction in general acts to minimize the discrepancies in the measured dynamical mass with respect to the intrinsic values. However, the result that applying the asymmetric drift correction does not change the measured dynamical mass for the idealised Narcissus observations shows that the presence of an asymmetric light distribution can lead to bias in the dynamical mass.

\noindent The literature shows that the physical assumptions inherent in dynamical mass calculations can lead to over- or under- estimating the mass of a system, as illustrated by \cite{Simons19} and \cite{Kohandel19} using simulations. 
Our results suggest that despite the indication from our tests that observations overestimate the velocity dispersion, the dynamical mass obtained after correcting for the asymmetric drift acts as a reasonable proxy for the total mass enclosed up to the observable radius, at least in the axisymmetric case as exemplified by Opuntia. It must be noted that the measurement of inclination angle would present a further challenge to the measurement of dynamical mass in observations of real galaxies. If we assume an error of $\pm 10$ degrees on our inclination angle of 60\degree, this introduces a further $+15\%/-10\%$ error on our measured dynamical masses according to the relation M$_{\text{dyn}}\mathrm{sin}^2(i) \sim$ constant.

\subsubsection{Estimating Dynamical Mass from Integrated Spectra}

We test the method employed by \cite{Kohandel19} of measuring dynamical mass from the full width at half maximum of the emission line, according to the equation:
\begin{equation}
    M_{\text{dyn}}^{\text{est}} = \left( \frac{\text{FWHM}^2}{\gamma^2\text{sin}^2(i)} \right) \left( \frac{R}{G} \right)\,,
\end{equation}
where $\theta$ is the inclination angle of the galaxy, R is the galactic radius, G is the gravitational constant, and we follow \cite{Capak15} in assuming a value of 1.32 for the parameter $\gamma$, a constant controlled by line profile and galaxy properties. This method yields the following dynamical mass estimates: $3.22^{+0.3}_{-0.3} \times 10^{10} \text{M}_{\odot}$ and $4.27^{+0.6}_{-0.6} \times 10^{10} \text{M}_{\odot}$ (Opuntia) and  $2.5^{+1.5}_{-1.5} \times 10^{10} \text{M}_{\odot}$ and $1.6^{+0.6}_{-0.6} \times 10^{10} \text{M}_{\odot}$ (Narcissus) in the idealised and mock NIRSpec case respectively.
Conversely to the other methods discussed in this section which underestimate the total mass within 1~kpc for these galaxies, estimating dynamical mass from the emission line overestimates the mass in the case of Opuntia (though Narcissus is consistent within 1$\sigma$). This effect is understood through the results presented in Section \ref{outflowsection}, in which we see that the information recovered from integrated spectra may be contaminated by effects such as outflows, with no clear way to separate out the two velocity components. 

\section{Summary and Conclusions}
\label{Conclusions}

\noindent In this work, we created idealised and realistic mock NIRSpec observations for two star-forming galaxies, Opuntia and Narcissus at $z = 6.1$ and $6.8$ respectively, from the {\sc serra} simulations. They were selected as case studies as both are disks with $v/\sigma \approx 10$ from [C{\sc ii}] kinematics, but they have different warm gas kinematics. For one galaxy (Narcissus), the $v/\sigma$ measured from H$\alpha$ is similar to that from [C{\sc ii}], while for the other (Opuntia) the H$\alpha$ $v/\sigma$ is a factor of 2 lower. 

\noindent We aim to assess how reliably we may expect to recover kinematic information from JWST/NIRSpec IFU observations, and whether we are able to identify galactic features such as rotating disks and outflowing gas using kinematics at such redshifts. The comparison of the idealised and realistic mock NIRSpec/IFU observations provides insight into the extent to which the spectral and spatial resolution and the S/N ratio impact the measured kinematic parameters, while comparing kinematic measurements from both sets of mock observations with those from the intrinsic H$\alpha$ simulations allows us to examine the validity of the implicit assumptions made when performing kinematic fitting. 

Our main conclusions are summarised below. 
\begin{itemize}
    \item At $z \geq 6$, we may expect to be able to recover complex structure from disk galaxies having a contribution from gas in non-circular motions, or strong asymmetries, at the resolution and SNR typical of NIRSpec/IFU observations, through discrepancies between the data and a rotating disk model. 
    \item Non-circular motions can be robustly identified by leveraging the information contained in the data cubes and the PV diagrams.
    \item The presence of non-circular motions and asymmetries contribute significant biases to measurements of turbulence and hence rotational support (Figure \ref{fig:vsigma}). We see that for Narcissus, which has a strong axial asymmetry, the assumption of an axisymmetric disk leads to a factor of $\sim 3$ inflation in the measured turbulence, irrespective of spatial and spectral resolution. For Opuntia, which is affected by non-circular motions, the lower spatial and spectral resolution available in mock NIRSpec observations biases the measured turbulence to $2.3 \times$ higher than its intrinsic value.
    \item The recovered dynamical mass appears to be robust for axisymmetric systems despite biases in $v_\text{rot}$ and $\sigma$, with the value measured from our mock NIRSpec observations being within 1$\sigma$ of the total baryonic+DM mass for Opuntia. Disk asymmetry presents a further challenge for dynamical mass recovery, with both the idealized and mock NIRSpec observations for Narcissus yielding values for dynamical mass that are $\sim2\times$ lower than the intrinsic total mass. 
    \item We present a caveat to the method of identifying outflows via the fitting of a double-Gaussian profile consisting of a narrow and broad component to the galaxy integrated spectrum by illustrating (Figure \ref{fig:outflow}) that the integrated spectrum of Opuntia, a galaxy that appears to host strong outflows, does not reproduce this profile, and its outflows are only recoverable through a 3-dimensional analysis. 
\end{itemize}

In conclusion, we show that at the spatial and spectral resolution and S/N ratio that may be expected for NIRSpec/IFU observations of individual galaxies at $z > 6$, disk galaxies with strongly asymmetric structure, and those hosting significant outflows, are measurably distinct from the standard rotating disk model. For this to be apparent, we show that it is necessary to leverage the entire three-dimensional spatial and spectral information contained within the IFU data, rather than collapsing those data into 2-dimensional maps or inspecting 1-dimensional spectra.  We stress the importance of comparing the different kinematic results obtained from H$\alpha$ and [C{\sc ii}], using synergistic warm and cold gas observations in building characterisations of galaxy kinematics. 

\section*{Acknowledgements}
\noindent We thank the anonymous referee for their highly constructive input. 
SP acknowledges a doctoral studentship
awarded by the Science and Technology Facilities Coucil (STFC).
RS acknowledges an STFC Ernest Rutherford Fellowship (ST/S004831/1).
We acknowledge the award of the Leon Rosenfeld grant 121531 enabling this collaboration.
We acknowledge the CINECA award under the ISCRA initiative for the availability of high-performance computing resources and support from the Class B project SERRA HP10BPUZ8F (PI: Pallottini). 
We acknowledge the use of the computational resources of the Center for High-Performance Computing (CHPC) at the Scuola Normale Superiore, Pisa. 
This research makes use of the Python programming language \citep{python}, Astropy \citep{Astropy}, Matplotlib \citep{matplotlib}, NumPy \citep{numpy}, and SciPy \citep{scipy}.

\section*{Data Availability}
\noindent The derived data created in this research will be shared upon reasonable request to the corresponding author.

\newpage



\bibliographystyle{mnras}
\bibliography{lessons_learned_main}



\newpage

\appendix

\section{The Effect of Masking on Velocity Map Analysis}
\label{appendix-ganifs}

\begin{figure}
	\includegraphics[width=0.5\textwidth]{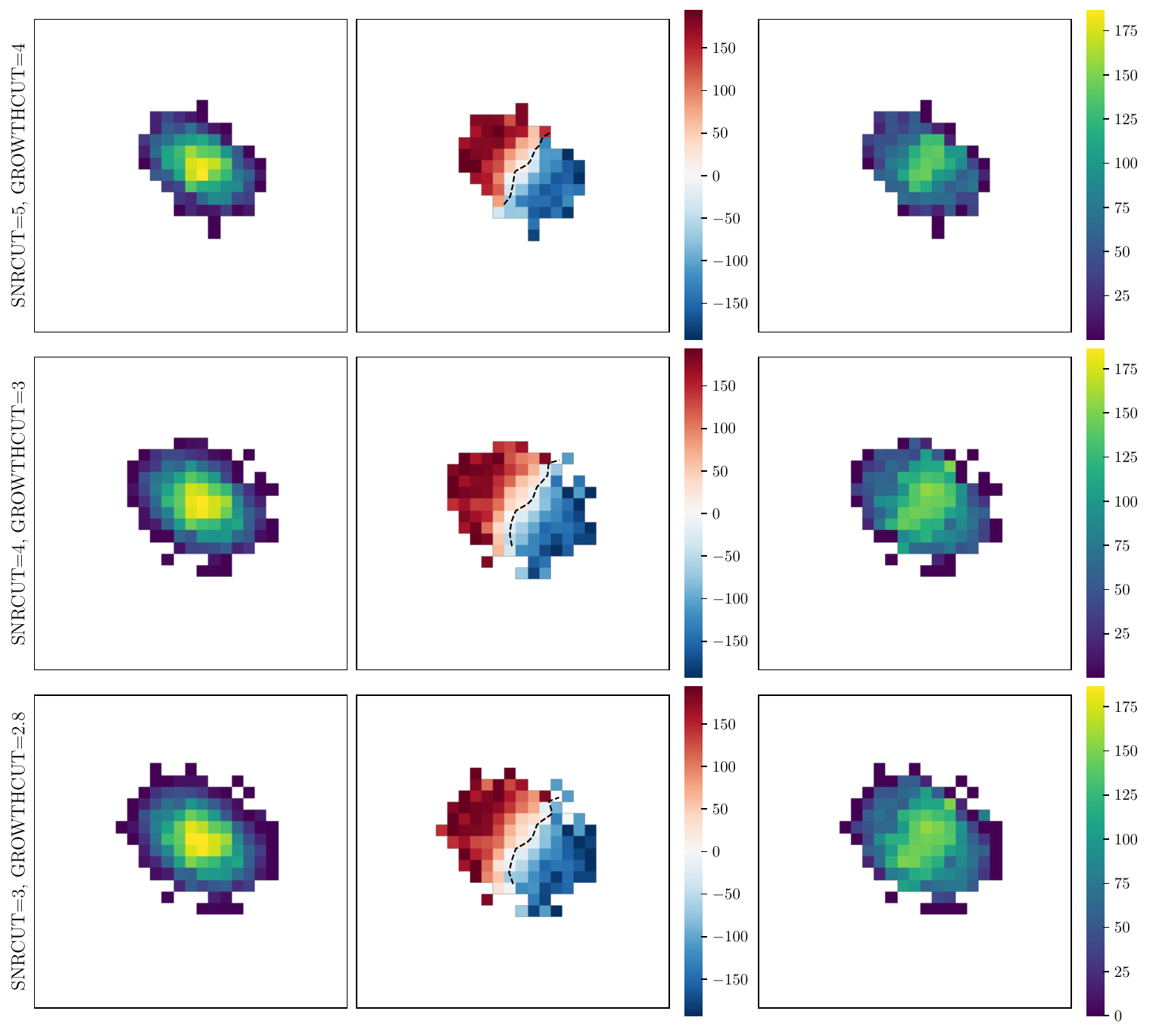}
    \caption{Moment maps for a symmetric disk model, containing no non-circular motions, with SNR=5, created using \textsuperscript{3D}\textsc{Barolo} with GROWTHCUT=4, 3, 2.8.}
    \label{fig:opuntiaganifs}
\end{figure}

\noindent We created model galaxies with \textsuperscript{3D}\textsc{Barolo} using the same parameters as the best-fit models for our galaxies. These models are by definition galaxy disks containing only circular motions. We convolve these model disk galaxies with the mock NIRSpec PSF, and add noise resulting in SNR=5, using these models to create velocity maps. In Figure \ref{fig:opuntiaganifs}, an example created using \textsuperscript{3D}\textsc{Barolo} is shown for a model disk galaxy to which we apply a selection of different masking restrictions. SNRCUT sets the SNR threshold of the moment map, while GROWTHCUT is a secondary SNR cut used to grow the mask \citep[see][for further details]{Barolo15}. We apply the masking combinations of (a) SNRCUT $=5$, GROWTHCUT $= 4$, (b) SNRCUT $=4$, GROWTHCUT $= 3$ and (c) SNRCUT $=3$, GROWTHCUT $= 2.8$. 
Although the underlying model contains only circular motions, the resulting $v=0$ contour shows deviations from a straight line.  with its shape changing depending on the mask that is applied. The exact shape of these deviations depends on the masking strategy adopted. This demonstrates that noise and masking alone can induce apparent distortions of the $v = 0$ isoline, even in the absence of non-circular motions. Therefore, such features should not be automatically interpreted as signatures of non-circular motions in real data.

\section{Channel Maps}
\label{channelmaps}
\begin{figure*}
	\includegraphics[width=0.9\textwidth]{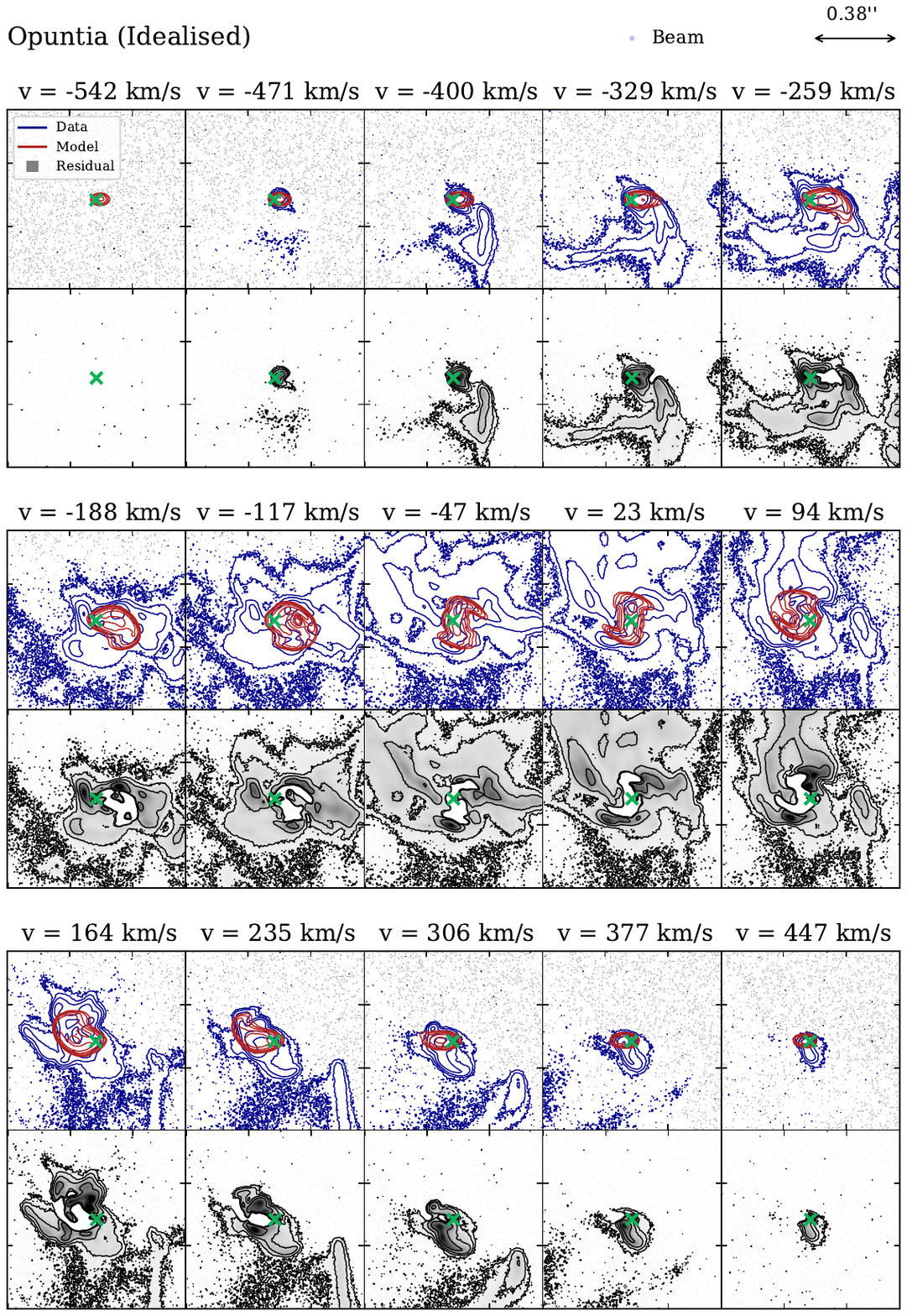}
    \caption{Representative channel maps for the idealised Opuntia observations (blue contours) and the corresponding model (red contours) alongside the residual (gray with black contours) as fitted by \textsuperscript{3D}\textsc{Barolo}. Contours are spaced at $3^n \times \sigma_\text{RMS}$ starting at n=1, where $\sigma_\text{RMS}$ is the standard deviation of pure noise channels. The green cross marks the fitted centre of the galaxy.}
    \label{fig:opuntia_ideal_channels}
\end{figure*}

\begin{figure*}
	\includegraphics[width=0.9\textwidth]{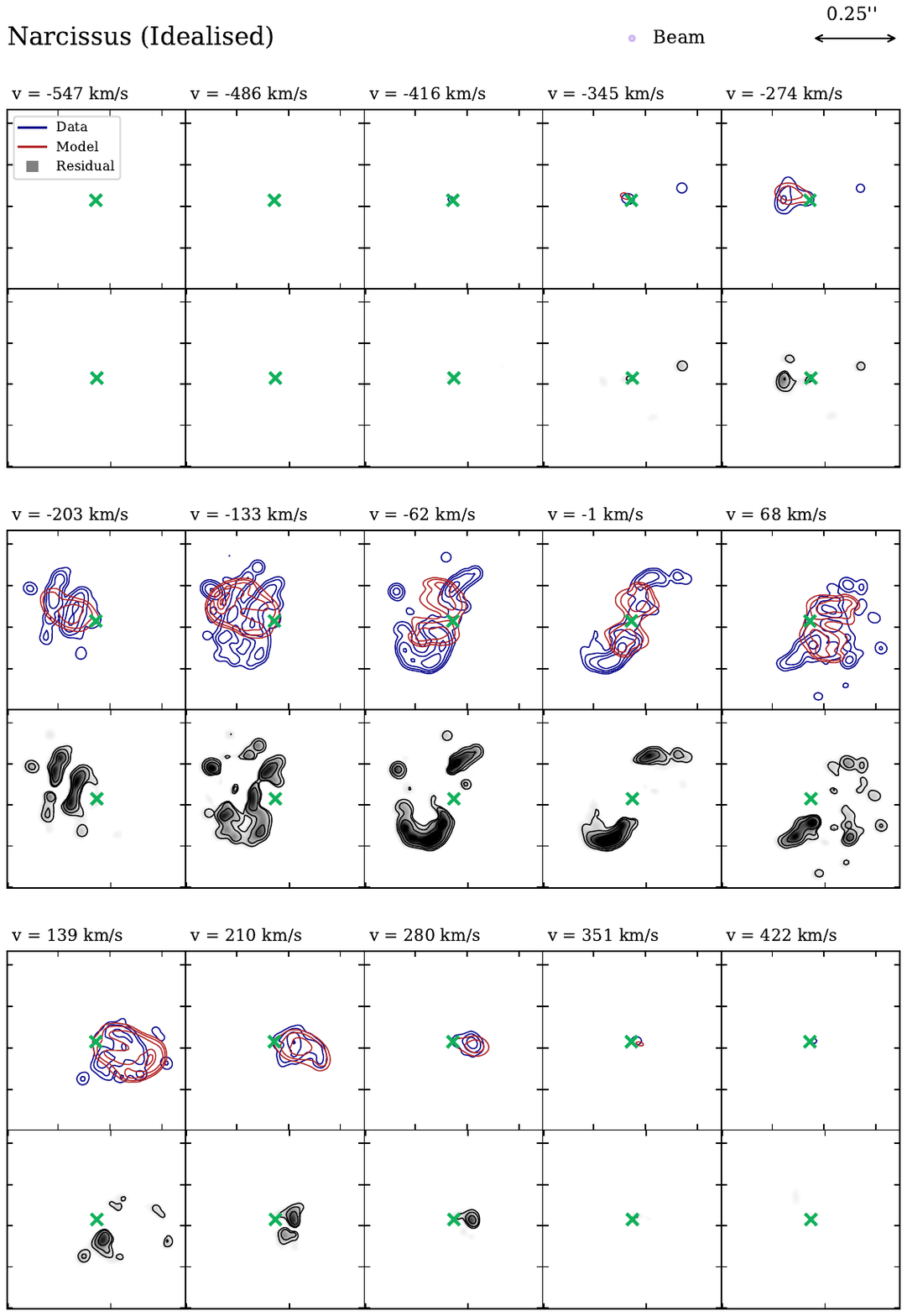}
    \caption{Channel maps with contour levels $3^n \times \sigma_\text{RMS}$ for the idealised Narcissus observations.}
    \label{fig:narcissus_ideal_channels}
\end{figure*}

\begin{figure*}
	\includegraphics[width=0.9\textwidth]{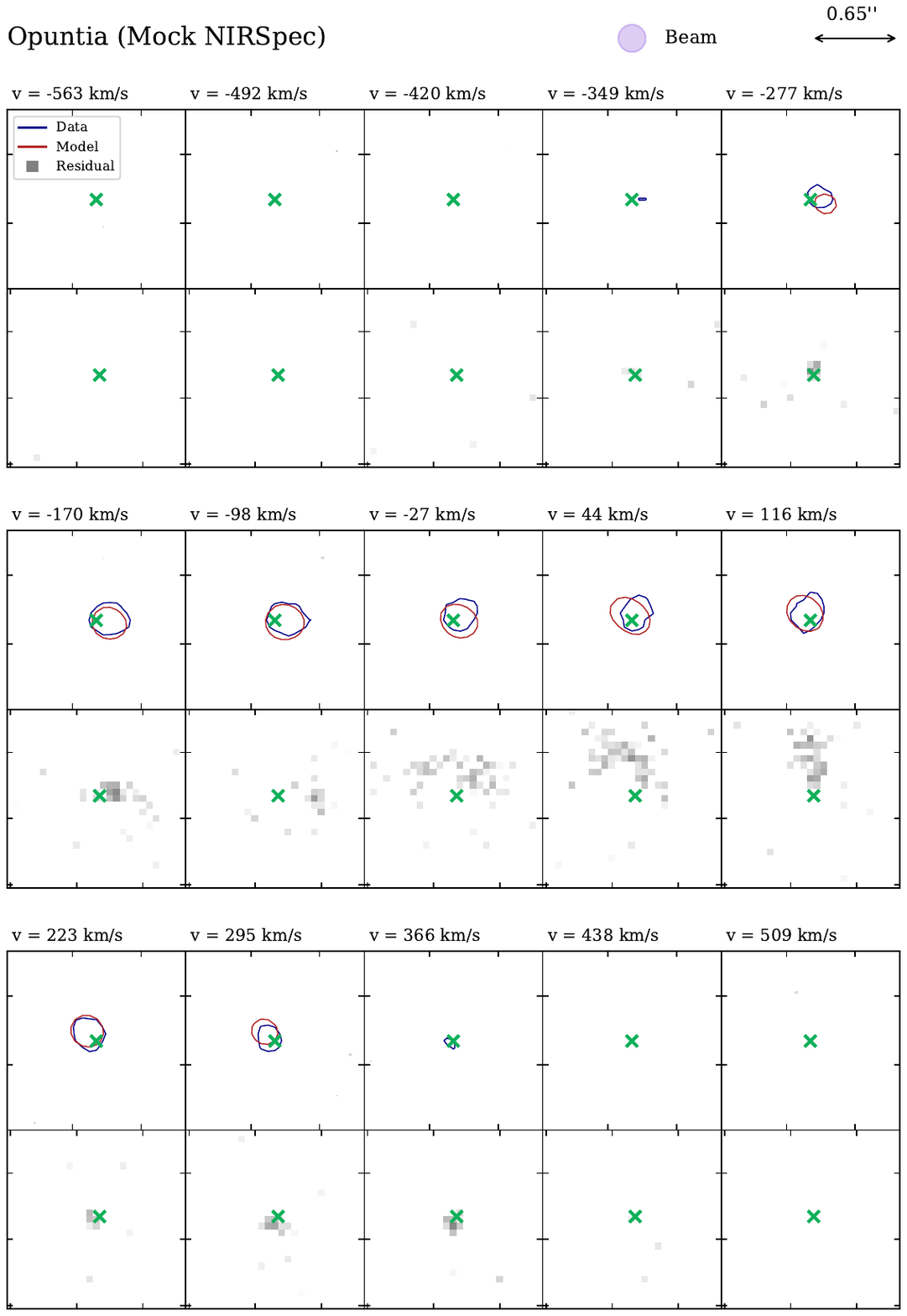}
    \caption{Channel maps with contour levels $3^n \times \sigma_\text{RMS}$ for the mock NIRSpec Opuntia observations.}
    \label{fig:opuntia_real_channels}
\end{figure*}

\begin{figure*}
	\includegraphics[width=0.9\textwidth]{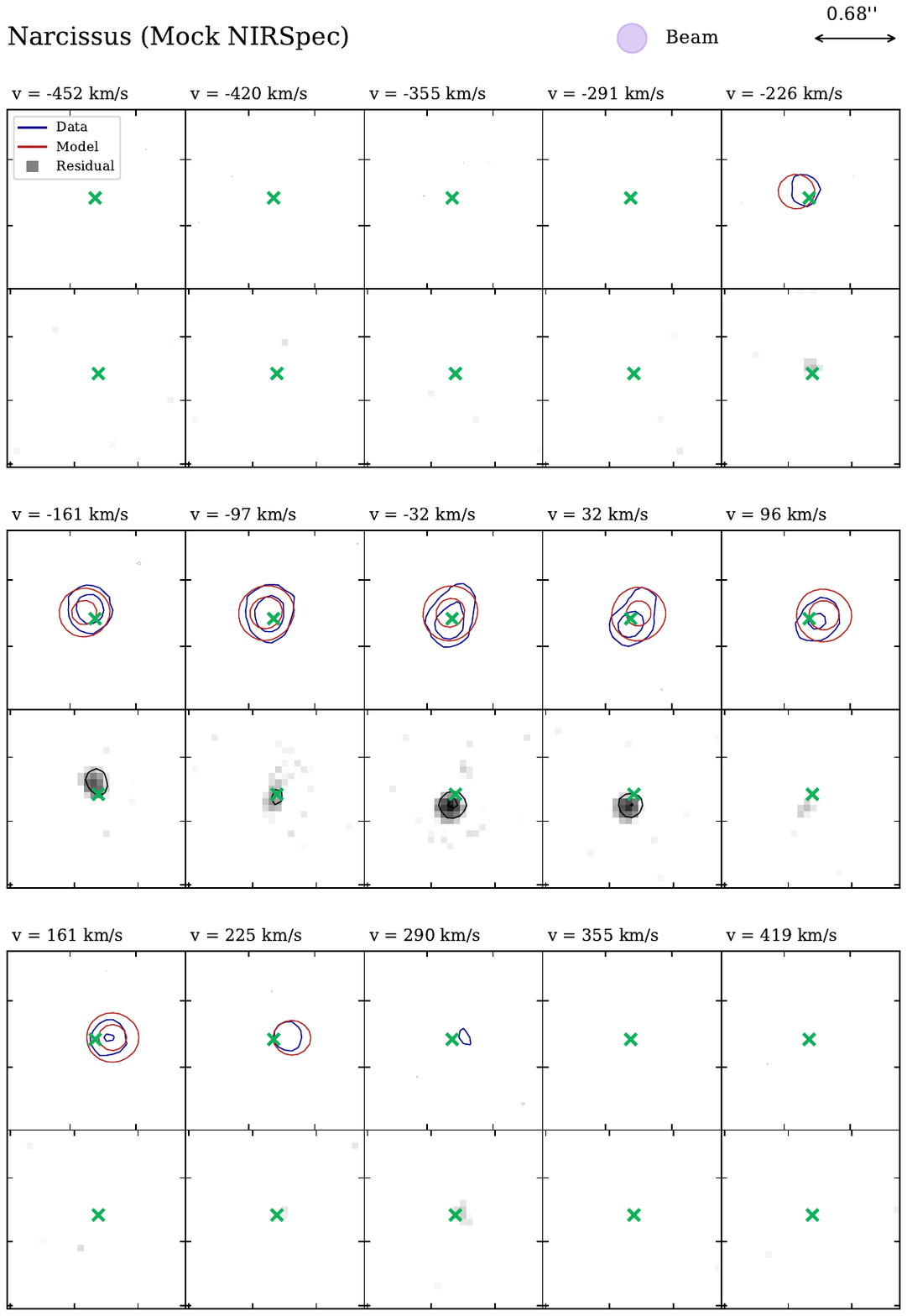}
    \caption{Channel maps with contour levels $3^n \times \sigma_\text{RMS}$ for the mock NIRSpec Narcissus observations.}
    \label{fig:narcissus_real_channels}
\end{figure*}

\noindent Figures \ref{fig:opuntia_ideal_channels}-\ref{fig:narcissus_real_channels} show channel maps of the H$\alpha$ emission for the idealised and mock NIRSpec observations of both galaxies, alongside the corresponding channel maps of the disk model fitted by \textsuperscript{3D}\textsc{Barolo}.

\section{Resolving Galaxies at $z > 6$}
\label{appendix_resolution}
\begin{table}
	\centering
	\caption{The effect of changing the SNR of an observation on the recovered kinematic measurements.}
	\label{tab:alt-snr}
	\begin{tabular}{|c|c|c|c|} 
		\hline
		Name & $v_{rot, max}$ & $\sigma$  & $v/\sigma$ \\
            \hline
		Opuntia (SNR=3) &  $290_{-12}^{+15}$ & $139_{-16}^{+13}$ & $2.1_{-0.6}^{+0.8}$ \\
		Narcissus (SNR=3) &  $189_{-19}^{+15}$ & $64_{-10}^{+10}$ & $3_{-0.7}^{+1.5}$ \\
        Opuntia (SNR=5, reference) & $290_{-28}^{+25}$ & $136^{+10}_{-14}$ & $2.1^{+0.2}_{-0.3}$\\
        Narcissus (SNR=5, reference)  & $193^{+14}_{-17}$ & $62^{+10}_{-11}$ & $3^{+0.9}_{-0.6}$ \\
		Opuntia (SNR=10) & $298_{-9}^{+8}$ & $138_{-14}^{+12}$ & $2.2_{-0.6}^{+0.7}$ \\
		Narcissus (SNR=10) & $193_{-14}^{+12}$ & $58_{-9}^{+9}$ & $3.3_{-0.9}^{+1.5}$  \\
		\hline
	\end{tabular}
\end{table}
\noindent 
We test the effect of changing the SNR on the kinematic properties recovered from our mock NIRSpec observations. To do this, we fit the same model with two tilted rings to different realisations of the mock NIRSpec data in which the SNR, as measured in a PSF-sized aperture at the outer edge of the galaxy, has been altered from $\sim 5$ to $\sim 3$ and $\sim 10$. The results of this test are shown in Table \ref{tab:alt-snr} along with a reproduction of the results at SNR $\sim 5$ for reference. We interpret the errors being smaller for Opuntia at lower SNR as resulting from the diffuse
component being less visible, making the model fit more straightforward. Correspondingly, at higher SNR the complex internal structure in Narcissus is more significant, which increases the uncertainty in the fitted rotating disk model. \\
\noindent At SNR $\sim 10$, the increased SNR enables us to fit an additional ring with \textsuperscript{3D}\textsc{Barolo}, and so we conduct a further test of factors affecting our measured kinematic properties, this time from increasing the number of independent resolution elements used for fitting from two to three. The v/$\sigma$ measurements of the realistic mock NIRSpec observations are then $2.2_{-0.2}^{+0.2}$ for Opuntia, an increase within 1$\sigma$, and $3.9_{-0.6}^{+0.7}$ for Narcissus, a increase within 1$\sigma$. The minor difference indicates that increasing the number of resolution elements is not effective for improving kinematic measurements without any corresponding improvement in spatial resolution.

\section{Comparison of Disk and Diffuse Emission} 
\label{separated_emission_components}

\begin{figure}
	\includegraphics[width=\columnwidth]{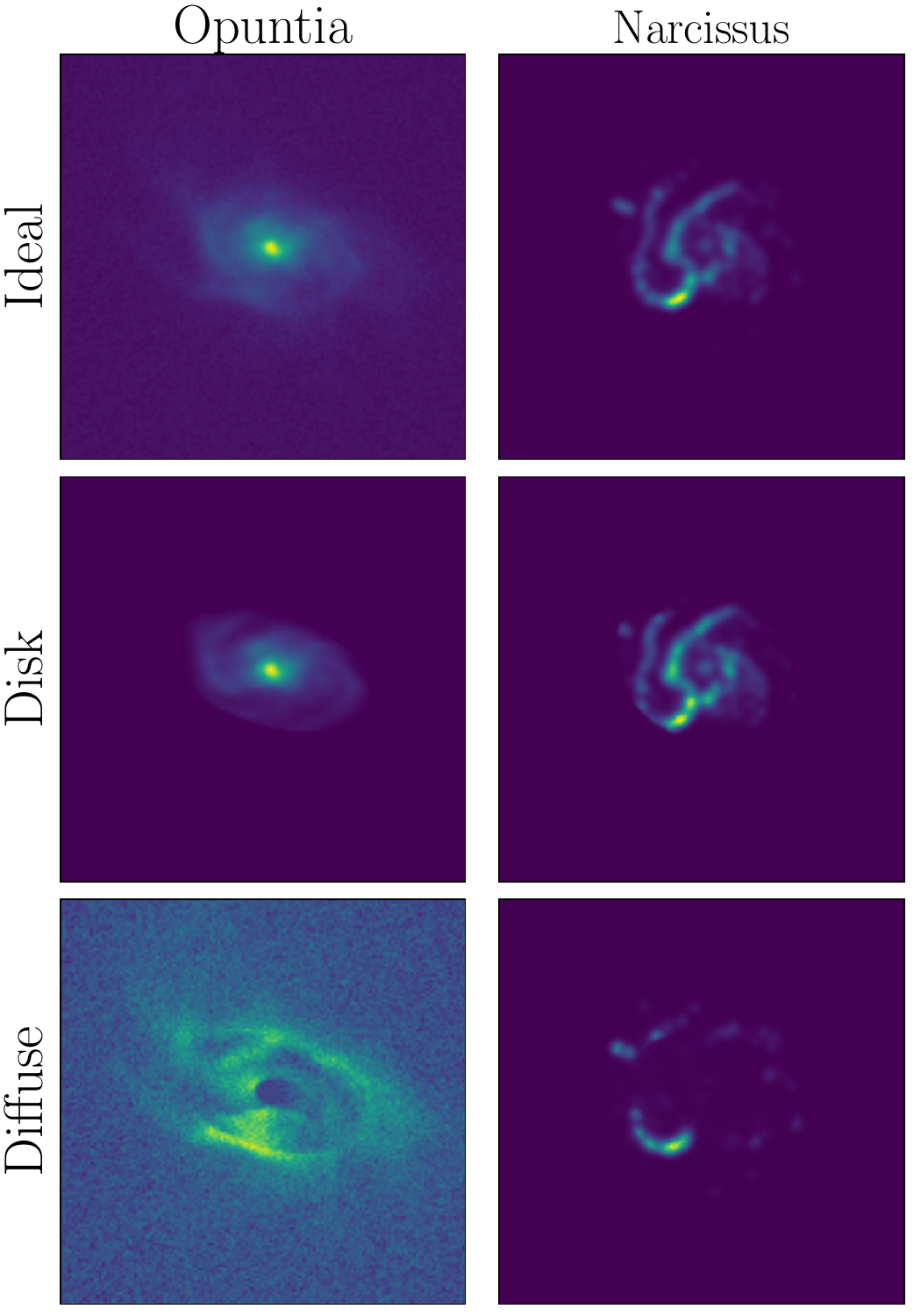}
    \caption{Moment-0 diagrams of the total idealised galaxy emission (first row), the disk component (second row) and the diffuse component (third row).}
    \label{fig:diskdiffusemom0}
\end{figure}

\noindent Figure \ref{fig:diskdiffusemom0} shows the moment-0 map of emission from the idealised galaxy observations, and the disk and diffuse components isolated by masking. 

\section{Double Gaussian Fitting to Mock NIRSpec Spectra}
\label{doubleGaussian}

\begin{figure}
	\includegraphics[width=0.5\textwidth]{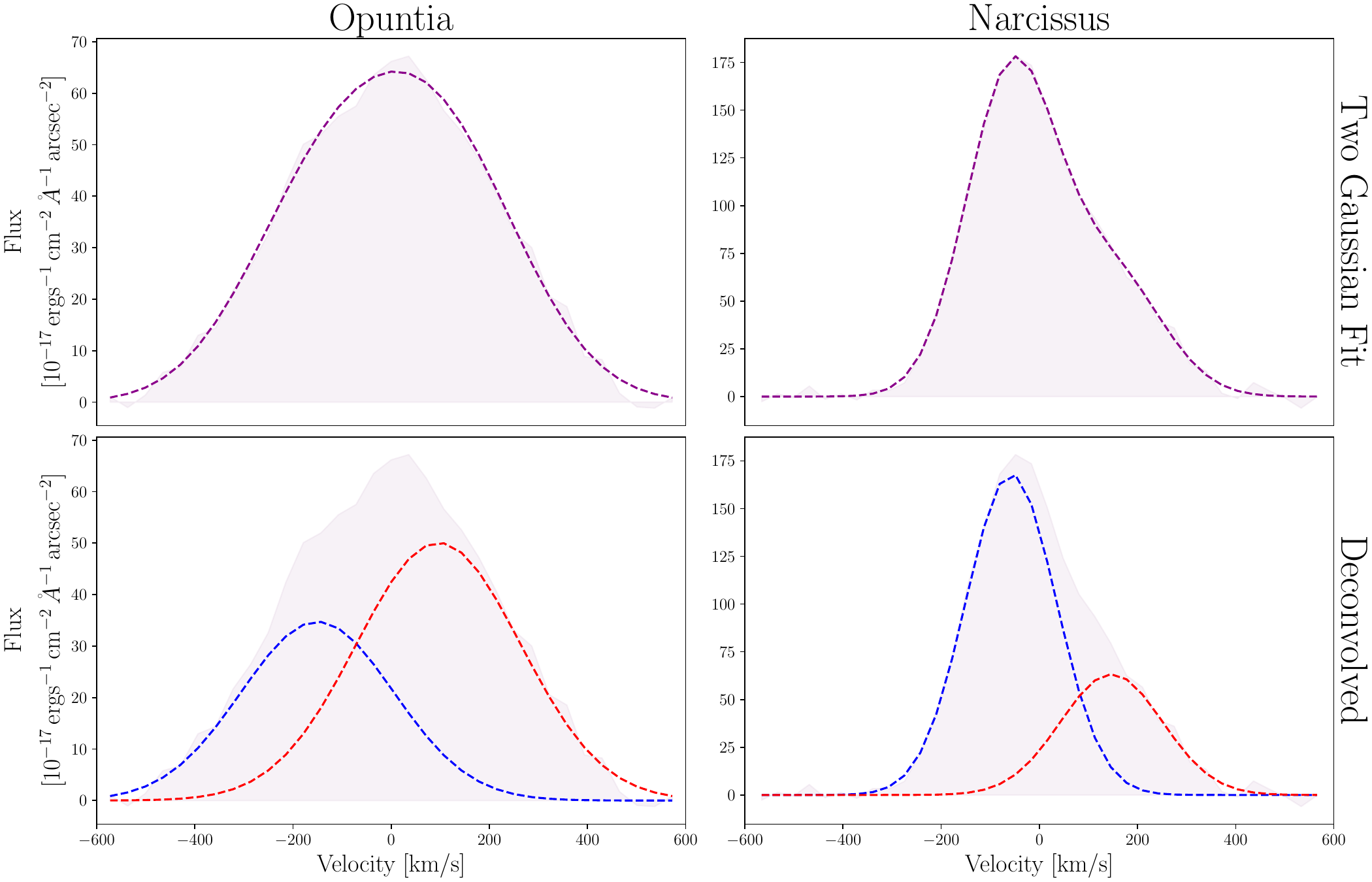}
    \caption{Integrated spectra of the Opuntia and Narcissus mock NIRSpec observations, fitted with a composite model (top row) consisting of two Gaussian components (second row). }
    \label{fig:dgfit}
\end{figure}

\noindent Figure \ref{fig:dgfit} shows a composite double-Gaussian model fitted to the integrated spectra of the realistic mock NIRSpec observations. This method is a widely employed technique to identify outflows from spectra, particularly where 3D data is not available; if the integrated spectrum is best fit by narrow and broad Gaussian peaks, this is considered evidence of outflows (see discussion in Section \ref{outflowsection}). This method would not suggest the presence of outflows in these mock NIRSpec observations, despite non-circular gas contributing significantly to the emission of Opuntia.

\section{Dynamical Mass Calculation}
\label{appendix_mdyn}
\noindent   To evaluate the asymmetric drift correction for the realistic mock NIRSpec observations, we assume that the first term in equation \ref{eqn_asymdrift} goes to zero, as it would not be meaningful to analytically evaluate the relation $\partial ln(\sigma)/\partial R$ when we were able to fit only two rings and so have only two data points. Thus, the equation reduces to:$$ v_A^2 = \frac{R\sigma^2}{R_\text{d, gas}} $$We measure $R_\text{d, gas}$ by fitting the moment-0 map with a S\'{e}rsic profile to determine the effective radius, and then converting this to the scale radius according to $R_\text{eff} = 1.68R_d$. We employ the \texttt{Sersic2D} function from the \texttt{astropy} modeling library to create the model, which we convolve with the same PSF as the data using the \texttt{PetroFit} routine \texttt{PSFConvolvedModel2D}. For both galaxies we fix the S\'{e}rsic index $n=1$, assuming an exponential disk, and for Narcissus we additionally fix the central co-ordinates to avoid the fit being skewed by the galaxy's asymmetries.

\label{lastpage}
\end{document}